\documentclass[sigconf]{acmart}

\acmJournal{PACMPL}
\AtBeginDocument{%
  }

\setcopyright{acmlicensed}
\copyrightyear{2018}
\acmYear{2018}
\acmDOI{XXXXXXX.XXXXXXX}

\usepackage[utf8]{inputenc} 
\usepackage{amsfonts}       
\usepackage{multirow}       
\usepackage{nicefrac}       

\usepackage{makecell}
\usepackage{enumitem}
\usepackage{amsmath} 
\usepackage{bm}
\usepackage[ruled,vlined,linesnumbered,noresetcount]{algorithm2e}
\SetAlgoSkip{smallskip}
\SetAlgoInsideSkip{noSkip}
\usepackage{subcaption}

\begin{document}


\title{Search-Based Multi-Trajectory Refinement for Safe C-to-Rust Translation with Large Language Models}


\author{HoHyun Sim}
\email{tlaghgus0425@korea.ac.kr}
\affiliation{%
  \institution{Korea University}
  \city{Sejong}
  \country{South Korea}
}

\author{Hyeonjoong Cho}
\authornote{Corresponding author.}
\email{raycho@korea.ac.kr}
\affiliation{%
  \institution{Korea University}
  \city{Sejong}
  \country{South Korea}}

\author{Yeonghyeon Go}
\email{doori74126@korea.ac.kr}
\affiliation{%
  \institution{Korea University}
  \city{Sejong}
  \country{South Korea}
}

\author{Sadegh AlMahdi Kazemi Zarkouei}
\email{skazemiz@CougarNet.UH.EDU}
\affiliation{%
  \institution{University of Houston}
  \city{Houston}
  \state{TX}
  \country{USA}}

\author{Zhoulai Fu}
\email{zhoulai.fu@sunykorea.ac.kr}
\affiliation{%
 \institution{SUNY Korea}
 \city{Incheon}
 \country{South Korea}}

\author{Ali Shokri}
\email{ashokri@Central.UH.EDU}
\affiliation{%
  \institution{University of Houston}
  \city{Houston}
  \state{TX}
  \country{USA}}

\author{Binoy Ravindran}
\email{binoy@vt.edu}
\affiliation{%
  \institution{Virginia Tech.}
  \city{Blacksburg}
  \state{VA}
  \country{USA}}

\renewcommand{\shortauthors}{Trovato et al.}

\begin{abstract}
The C programming language has been foundational in building system-level software. However, its manual memory management model frequently leads to memory safety issues. In response, a modern system programming language, Rust, has emerged as a memory-safe alternative. Moreover, automating the C-to-Rust translation empowered by the rapid advancements of the generative capabilities of LLMs is gaining growing interest for large volumes of legacy C code. Leveraging the LLM capabilities for the C-to-Rust translation introduces distinct challenges, unlike the math or commonsense QA domains where the LLM capabilities have been predominantly applied. First, the scarcity of parallel C-to-Rust datasets hinders the retrieval of suitable code translation exemplars for in-context learning. Second, unlike math or commonsense QA problems, the intermediate steps required for C-to-Rust are not well-defined. Third, it remains unclear how to organize and cascade these intermediate steps to construct a correct translation trajectory. While existing LLM-based approaches have achieved some success, they have relied on iterative code refinement along a single search trajectory on a C-to-Rust problem space and have not explored the use of systematic search mechanisms to navigate the space of possible refinement trajectories. To address these challenges in the C-to-Rust translation, we propose the MCTS (Monte Carlo Tree Search)-Guided LLM refinement technique for automated C-to-safe-Rust translation (LAC2R). LAC2R uses MCTS to systematically explore multiple refinement trajectories and organize the LLM-induced intermediate steps for correct translation. We experimentally demonstrated that LAC2R effectively conducts C-to-Rust translation on large-scale, real-world benchmarks. On GNU coreutils benchmarks (7 programs, \textless{}15 KLoC), LAC2R achieves an average function-compile rate (FCR) of 97\%, surpassing two major counterparts, such as 
C2SaferRust (74--80\%) and EvoC2Rust (59\%); EvoC2Rust, however, passes 0\% of behavioral test cases, yielding functionally inequivalent translations due to brace mismatches introduced by LLM code generation in macro-heavy functions. On Laertes benchmarks (10 programs, \textless{}96 KLoC), LAC2R achieves an average FCR of 88\% with full project compilability, while EvoC2Rust, despite reporting the highest average safety ratio (0.75), fails to produce a project-compilable translation across all ten benchmarks. Among approaches that yield valid, executable translations, LAC2R achieves the best safety ratio on both suites (0.50 vs. 0.32--0.33 for coreutils; 0.54 vs. 0.26--0.29 for Laertes) and reduces linter warnings by 25--39\% compared to C2SaferRust. On small-scale benchmarks, the TRACTOR Public-Tests (Battery-01 + \texttt{P00\_perlin\_noise}), LAC2R is the only method that simultaneously attains the highest safety ratio (0.95 vs. 0--0.88), perfect project-level correctness (100\% program compilation and 100\% program pass rate), and the fewest linter warnings among the compared methods. The results indicate that LAC2R substantially improves safety and reliability in program-level C-to-Rust translation.
\end{abstract}

\ccsdesc[500]{Computing methodologies~Artificial intelligence}
\ccsdesc[500]{Software and its engineering~General programming languages}

\keywords{Monte Carlo Tree Search, Large Language Model, C-to-Rust}

\received{20 February 2007}
\received[revised]{12 March 2009}
\received[accepted]{5 June 2009}

\maketitle

\section{Introduction}

The C programming language has been foundational for building system-level software, such as operating systems, embedded systems, and performance-critical applications. Its fine-grained control over hardware makes it indispensable in such domains. However, C’s manual memory management model frequently leads to memory safety issues such as buffer overflows, dangling pointers, and data races. Industry reports have estimated that 70\% of their security vulnerabilities stem from these
memory safety issues~\cite{stoep2019queuethehardening, msrc2019weneedasafersystems} and the US government recently emphasized the importance of transitioning to safe programming languages~\cite{thewhitehouse2024backtothebuildingblocks, thewhitehouse2024nationalcybersecuritystrategy}. In response, a modern system programming language, Rust, has emerged as an alternative that offers memory safety by enforcing a strict ownership and borrowing model at compile time. Rust has been successfully adopted in several projects, including Mozilla Firefox and AWS Firecracker. However, a large number of legacy C codes still exist and manually converting them into Rust requires significant cost, which has driven growing interest in automating the C-to-Rust translation.

Existing automatic C-to-Rust translation techniques are generally classified into two categories: rule-based and LLM-based approaches. Traditional rule-based approaches, such as C2Rust~\cite{c2rust_manual}, aim to preserve equivalence during translation. However, they often produce non-idiomatic Rust code that contains unsafe blocks and low-level constructs, which undermines both safety and maintainability. In contrast, LLM-based approaches can generate idiomatic and safer Rust code, as LLMs are trained on the corpora of human-written code. Nevertheless, they lack equivalence guarantees due to the hallucination problem inherent to LLMs. To address this, recent approaches combine the generative capabilities of LLMs and the verifiable determinism of external tools, such as code analyzers and validators, to mitigate hallucinations.

\begin{figure}[hbpt]
    \centering  
    \includegraphics[width=\linewidth]{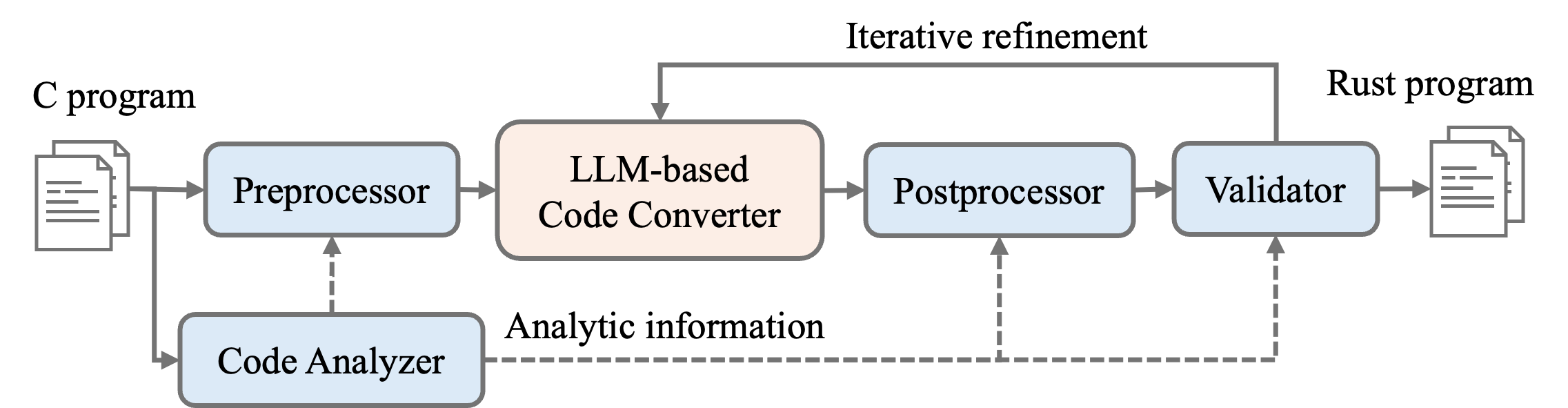} 
    \Description{Decorative image; no additional information.}
    \caption{The common execution flow of existing LLM-based C-to-Rust translation methods. The dotted connections are selectively established. }
    \label{fig:exe_flow}
\end{figure}

Several LLM-based approaches have been proposed recently, each with unique features, as listed in Table~\ref{tbl:llm-approaches-structure}. At a high-level, however, they share a similar execution flow as illustrated in Figure~\ref{fig:exe_flow}. In this flow, a preprocessor decomposes the original C code into small snippets and, if necessary, translates them into initial Rust code. In parallel, a code analyzer extracts additional information about the original code and provides it to the LLM-based code converter, validator, or postprocessor, depending on the design choice. When the information is provided in the form of the prompt, the LLM converter generates a Rust code snippet and the postprocessor prepares it for validation. Then, the validator checks the correctness of the generated code in several dimensions, such as compilability and equivalence. Based on feedback from the validator, the LLM converter iteratively refines the Rust code until predefined termination conditions are satisfied. This iterative code refinement strategy using the external feedback has been successfully applied to several benchmarks as shown in Table~\ref{tbl:llm-approaches-experiment}.

\noindent \textbf{Challenge.} Leveraging the LLM capabilities for the C-to-Rust translation introduces distinct challenges, unlike the math or commonsense QA domains where the LLM capabilities have been predominantly applied~\cite{10.1145/3719341}. First, the scarcity of parallel C-to-Rust datasets makes it difficult to retrieve suitable translation exemplars for in-context learning, which limits the effectiveness of existing multi-step prompting techniques such as few-shot chain-of-thought reasoning~\cite{10.5555/3600270.3602070}. Second, unlike math or commonsense QA problems where intermediate steps can be logically defined and easily verified, the intermediate steps required for C-to-Rust are not well-defined. Third, even if individual intermediate steps for C-to-Rust are given, it remains unclear how to organize a promising multi-step trajectory toward a correct C-to-Rust translation. As a result, addressing the C-to-Rust translation task requires methods that can induce LLMs to generate effective intermediate-step results that facilitate the C-to-Rust transformation, as well as systematic search mechanisms to coordinate those intermediate steps effectively.

However, existing LLM-based approaches have relied on iterative refinement along a single search trajectory, leaving the multi-trajectory search space unexplored despite its natural formulation as a sequential decision-making problem.

\noindent \textbf{Contributions.} Motivated by these challenges, we propose the \textit{MCTS-Guided LLM refinement for Automated C-to-safe-Rust translation} (LAC2R). To build a code translation technique capable of systematically planning the intermediate steps for C-to-Rust translation, we formulate the translation as a sequential decision-making problem in a code refinement search space. To generate and navigate the diverse reasoning trajectories for this task, we propose using the Monte Carlo Tree Search (MCTS)~\cite{6145622} as a framework where the intermediate reasoning steps, such as iterative code refinement, are systematically structured. MCTS enables a principled balance between exploration and exploitation by organizing a tree-like reasoning structure guided by the \textit{Upper Confidence bounds applied to Trees} (UCT) with reward evaluation. We propose a reward calculation tailored for C-to-Rust and assess its effectiveness. In addition, we leverage multiple heterogeneous LLMs to generate intermediate refinement steps for C-to-Rust code translation. Leveraging different LLMs enhances diversity in Rust refinement candidates as the heterogeneity of their training datasets encourages their complementary translation.

In addition, we experimentally show that LAC2R effectively conducts C-to-Rust translation on large-scale, real-world benchmarks compared to prior counterparts. On the GNU coreutils benchmarks, LAC2R achieves the highest average FCR (97\%), surpassing C2SaferRust (74--80\%) and EvoC2Rust (59\%); however, EvoC2Rust passes 0\% of behavioral test cases across all seven benchmarks, indicating that its translations do not preserve the original program behavior. On the Laertes benchmarks, LAC2R achieves an average FCR of 88\% while maintaining full project compilability; in contrast, EvoC2Rust, despite reporting the highest safety ratio (0.75), fails to produce a project-compilable translation in any of the ten benchmarks, rendering its safety improvement unverifiable in practice. Among approaches that produce executable, behaviorally equivalent translations, LAC2R achieves the highest safety ratio on both suites, 0.50 for coreutils (vs. 0.32--0.33 for C2SaferRust) and 0.54 for Laertes (vs. 0.26--0.29), and reduces linter warnings by 25--39\%. We also evaluate \textsc{LAC2R} on the \textsc{TRACTOR} Public-Tests benchmark (Battery-01 + \texttt{P00\_perlin\_noise)}. On this benchmark, LAC2R outperforms counterparts such as SACTOR, EvoC2Rust, and RustAssure across both safety and code quality metrics. Compared to C2SaferRust, LAC2R matches the highest project-level correctness (100\% program compilation and 100\% program pass rate) while simultaneously attaining a superior safety ratio and substantially fewer linter warnings, demonstrating that correctness and safety need not be traded off against one another.


\section{Proposed Approach}
\label{sec:approach}

\begin{figure}[hbpt]
    \centering  
    \includegraphics[width=\linewidth]{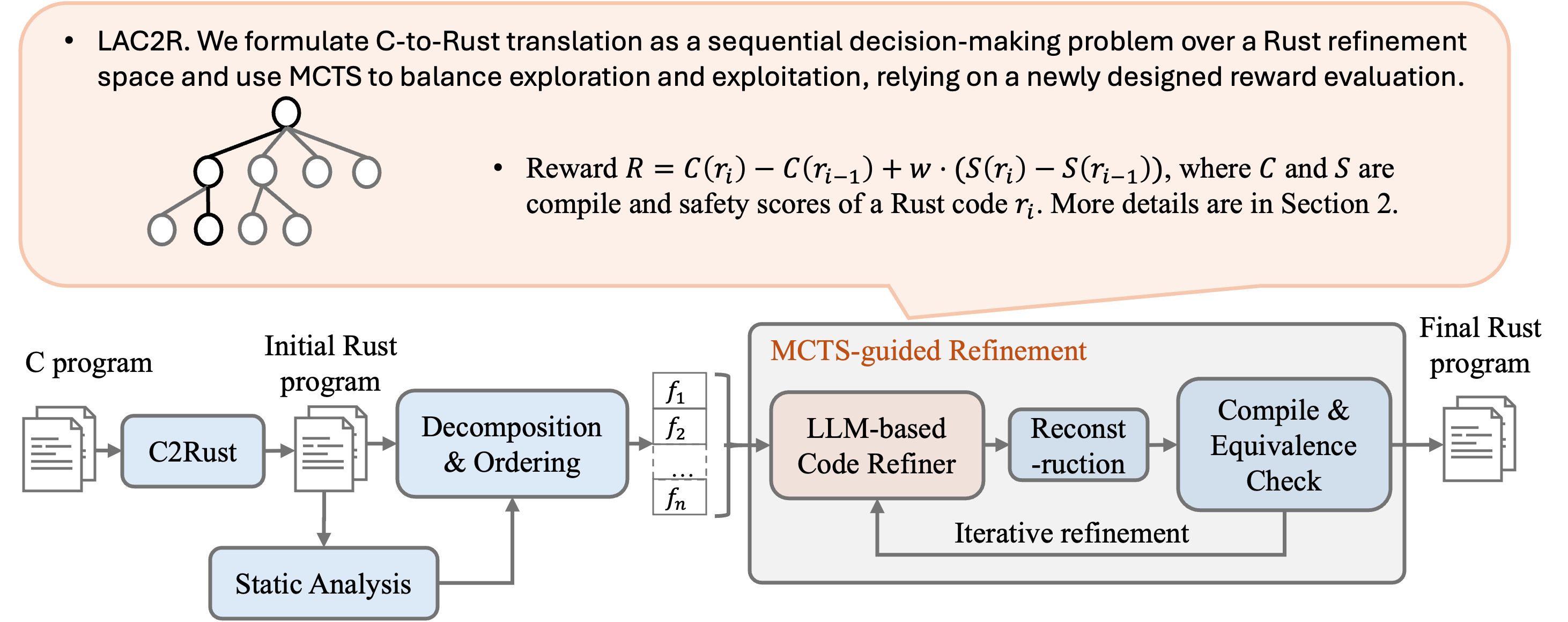} 
    \Description{Decorative image; no additional information.}
    \caption{Overview of LAC2R’s execution flow with an illustration of its key components.}
    \label{fig:overview}
\end{figure}

\subsection{Overview}
\label{sec:overview}

To address the large scale of legacy C programs and the restricted context length of LLMs, LAC2R takes an iterative function-wise translation strategy. As illustrated in Figure~\ref{fig:overview}, the execution starts with C2Rust that converts a target C program to an initial unsafe Rust. This initial compilable Rust program enables focused compile check during following function-wise replacement. Then, the Rust program is decomposed into functions and ordered according to a dependency graph constructed by a static analyzer. Using this information, LAC2R repeatedly refines the Rust functions in the predefined order to generate safer and behaviorally equivalent code. 

To refine a single Rust function, LAC2R searches for the best refinement trajectory from the initial unsafe Rust function to the final candidate using MCTS. MCTS constructs a search tree by repeatedly expanding edges until a termination criterion is met, where each expansion corresponds to one iteration comprising LLM-based code refinement, program reconstruction using the refined function, and validations. Since MCTS prioritizes the trajectories with higher validation scores, it progressively concentrates on the trajectories that produce safer and test-equivalent Rust code. 

\begin{figure}[hbpt]
    \centering  
    \includegraphics[width=\linewidth]{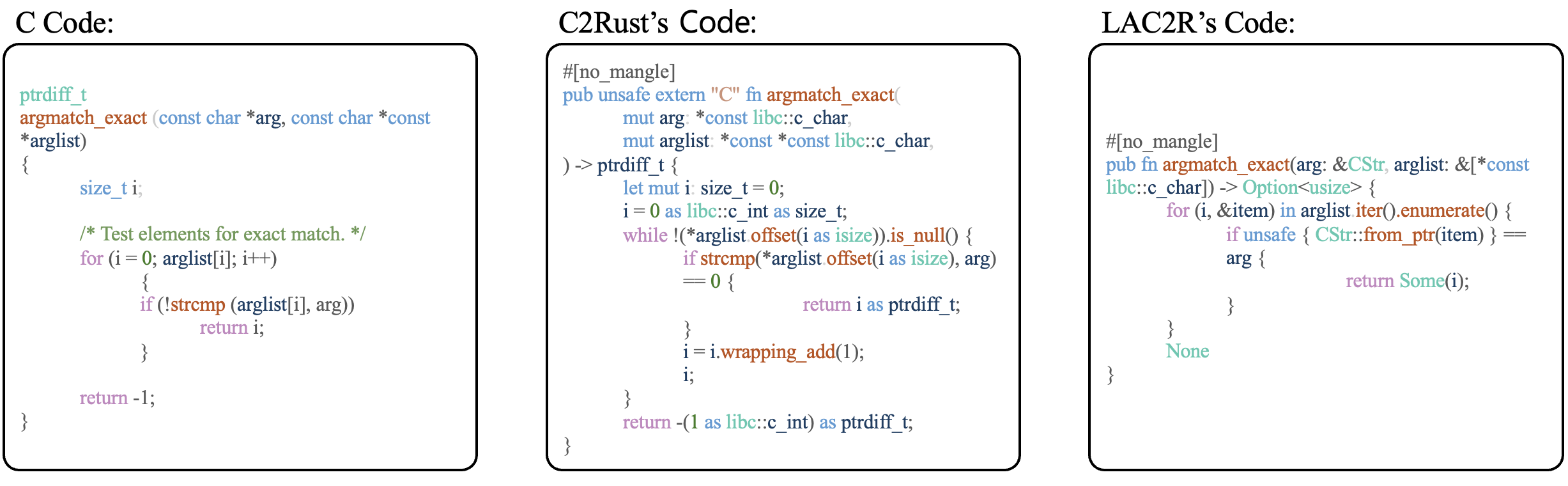} 
    \Description{Decorative image; no additional information.}
    \caption{LAC2R's translation example including a target C function, its unsafe Rust conversion by C2Rust, and its final Rust translation. \texttt{argmatch\_exact()} is a sub-function of \texttt{tail} benchmark in the GNU coreutils suite.}
    \label{fig:cexamples}
\end{figure}

Figure~\ref{fig:cexamples} shows an example of LAC2R's C-to-Rust translation. LAC2R first converts the target C function, \texttt{argmatch\_exact()}, to an unsafe Rust code using C2Rust. Then, LAC2R refines the unsafe Rust to the safe one relying on the MCTS framework. 

\subsection{Problem Formulation}
\newcommand{\postprocess}{\overset{\text{postprocess}}{\longleftarrow}}

We formulate the iterative Rust refinement of a single function as a finite-horizon, constrained sequential decision-making problem that aims at maximizing the safety of the Rust result. Formally, the problem is defined by a tuple $(\mathcal{R}, \mathcal{A}, \tau, S, c_0)$, where $\mathcal{R}$ denotes a potentially unbounded state space of Rust codes for the target function, $\mathcal{A}$ denotes a set of refinement actions, $\tau:\mathcal{R} \times \mathcal{A} \to \mathcal{R}$ denotes the state transition representing the code refinement in our context, $S:\mathcal{R}\to\mathbb{R}$ is a safety evaluation of Rust program, and $c_0$ is a target C code.

Given its C2Rust translation $r_0=\mathrm{C2Rust}(c_0)$, the objective is to obtain a refined Rust program $r_N$ after $N$ steps that is both safe and test-equivalent to $c_0$. We denote a fixed test suite by $\mathcal{T}_{\mathrm{test}}$. We let $\mathrm{Exec}(p, t)$ be the observable behavior, such as output or exit status, of program $p$ on test $t$, and write $p \equiv_{\mathcal{T}_{\mathrm{test}}} q$ iff $\mathrm{Exec}(p,t)=\mathrm{Exec}(q,t)$ for all $t\in\mathcal{T}_{\mathrm{test}}$.
We also use a validator $V:\mathcal{R}\to\mathcal{Z}$ that returns compile/test feedback signals.

We seek a sequence of actions $(a_1, ..., a_N)$ with $r_{i+1}=\tau(r_i,a_i)$, formalized as: 
\begin{equation}
\label{eq:objective}
\max_{(a_1,\dots,a_N)} \; S(r_N)
\quad\text{s.t.}\quad
r_N \equiv_{\mathcal{T}_{\mathrm{test}}} c_0.\;\; 
\end{equation}
A transition is induced by an LLM-based code refiner $F$ and a post-processing step $\Pi$:
\begin{equation}
\label{eq:transition}
\tau(r,a) \;=\; \Pi\!\left(F\!\left(\mathrm{prompt}_a\!\big(r, V(r)\big)\right)\right).
\end{equation}
where $prompt_a$ denotes a prompt prepared for an action $a$. To achieve the objective, we propose LAC2R employing MCTS to search for an optimal sequence of actions $(a_1, ..., a_N)$. 



\subsection{LAC2R Design}
\label{subsec:lac2r}

\textbf{MCTS.} LAC2R leverages MCTS to construct promising code refinement trajectories. MCTS builds a search tree where each node represents a state of intermediate Rust code and each edge represents a refinement transition, enabling principled exploration--exploitation balance via UCT. The detailed procedure is described in Algorithm~\ref{alg:mcts-translation}.


\noindent \textbf{Objective and Reward Functions.} LAC2R aims to maximize the safety function $S$ as defined in Equation~\ref{eq:objective}. We define $S$ to represent the reduction rate in five categories of unsafe Rust constructs, as introduced in the prior work~\cite{nitin2025c2saferrusttransformingcprojects}. Formally, the $S$ is defined as:
\begin{align}
\label{eq:safety}
&S(r_{i})=m(r_i) \cdot max\bigg\{ \bigg( 1-\frac{RPC_{i} + RPR_{i} + LUC_{i} + UCE_{i} + UTC_{i}}{RPC_{0} + RPR_{0} + LUC_{0} + UCE_{0} + UTC_{0}} \bigg),0 \bigg\},
\end{align}
where $RPC_i$, $RPR_i$, $LUC_i$, $UCE_i$, and $UTC_i$ denote the numbers of raw pointer declaration, raw pointer dereferences, lines of unsafe code, unsafe call expressions, and unsafe type casts in the Rust program with the function $r_i$, respectively. If the Rust program with $r_i$ is compilable, $m(r_i)=1$; otherwise, $m(r_i)=0$. The indicator $m(r_i)$ is adopted to prioritize compilability over safety and penalize non-compilable code.

The compile score $C(r_i)$ is computed as:
\begin{align}
C(r_i)=\frac{1}{|E_C(r_i)|+1},
\end{align}
where $|E_C|$ is the number of compile errors. 

To select a promising node for tree expansion, MCTS uses the UCT that is updated based on a reward. LAC2R defines the node reward $R$ based on both the compile score and the safety of the Rust code. Formally, the $R$ is computed as: 
\begin{align}
&R = R_V + w \cdot R_S= C(r_i)-C(r_{i-1})  + w \cdot (S(r_i)-S(r_{i-1})),
\end{align}

where $R_V$ denotes a reward computed using validation feedbacks and $R_S$ denotes a reward computed using safety metrics. The scalar $w$ is a weighting factor that balances the contributions of two reward terms. 

Since the reward is defined as the increment from each node to its child, the cumulative reward along a trajectory corresponds to the total improvement from the root to the terminal node. Thus, the MCTS policy of maximizing the expected accumulated reward steers the search toward both reducing compilation errors and increase the safety score. When compilation succeeds, the search effectively maximizes the objective function.

\noindent \textbf{Transition Action with Heterogeneous LLMs.} To obtain diverse intermediate refinements, LAC2R uses heterogeneous LLMs, whose distinct training datasets encourage complementary refinement effects.


LAC2R allows two types of actions for transitions, such as code refinement with validator feedback and without feedback, denoted by $(feedback, LLM_K)$ and $(null, LLM_k)$, respectively. $k\in \{1,...,K\}$ indexes available LLMs where $K$ is the total number. For the action with feedback, the LLM prompt incorporates feedback from validators, such as a compiler and a testcase executor for code refinement.

\begin{figure}[hbpt]
    \centering  
    \includegraphics[width=1.0\linewidth, height=2.1cm]{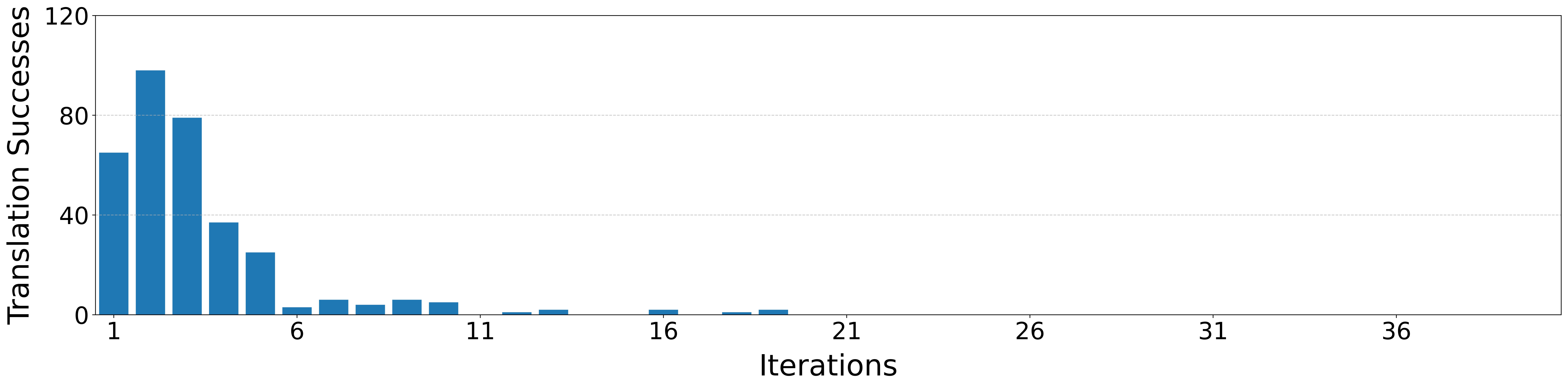} 
    \Description{Decorative image; no additional information.}
    \caption{The FRR distribution of \textsc{C2SaferRust}'s iterative Rust refinement over iterations.}
    \label{fig:success_rates}
\end{figure}


\noindent \textbf{Tree Configuration.} As shown in Table~\ref{tbl:llm-approaches-structure}, most existing LLM-based C-to-Rust translation methods, including \textsc{C2SaferRust}, rely on sequential, iterative code refinement based on external feedback. When running \textsc{C2SaferRust} on the seven coreutil benchmarks as in~\cite{nitin2025c2saferrusttransformingcprojects}, the distribution of function-replacement rate ($FRR$) over iterations is visualized in Figure~\ref{fig:success_rates}, where a function translation is counted as successful (i.e., $FRR$\,=\,1 for that function) when there are no compile-time or test-case execution errors. It indicates that more than 80\% of the successfully translated functions achieve this within the first five iterations. Additionally, iterative code refinement beyond five iterations tends to be less effective and no additional function is successfully translated after 20 iterations, although the total $FRR$ does not reach 100\%. To take advantage of the higher $FRR$ in the early iterations, LAC2R is designed to generate an increased number of initial Rust candidates. In addition, to improve the effectiveness of long iterative refinement, LAC2R incorporates diversity using heterogeneous LLMs. Moreover, to reduce the iteration length, we allow each node to spawn multiple child nodes as presented in~\cite{qi2024mutualreasoningmakessmaller}. 

\noindent \textbf{Algorithm.} LAC2R employs an MCTS-based search strategy as presented in Algorithm~\ref{alg:mcts-translation}. Each rollout performs four standard MCTS phases (Lines 6--10): UCT-guided selection to a promising leaf, expansion via \texttt{Expand} (Algorithm~\ref{alg:expand}), greedy simulation to a terminal node, and backpropagation of rewards along the full path. The process repeats for \texttt{num\_rollouts} iterations, and the final solution is selected from the constructed tree (Line 4).

Expansion, one of LAC2R's core phases, is described in detail in Algorithm~\ref{alg:expand}. When expanding the root (\textsc{Init}) node, multiple \textsc{Gen} children are created using heterogeneous LLMs (Lines 2--3), exploiting the higher success rates in the early iterations observed in Figure~\ref{fig:success_rates}. During non-root expansions, the LLM generates a candidate translation using the conversation history and any prior feedback, which is then inserted into the full program (Lines 4--5). If compilation succeeds, the node reward is computed based on the safety score and compile status (Line 7). When test suites are available, they are executed: if all tests pass, a \textsc{Success} child is added to the node (Line 10); otherwise, \textsc{Fix} children are added with test-execution errors as feedback (Line 11). If no test suites are available, a successful compilation alone is sufficient to add a \textsc{Success} child (Line 12). When compilation fails, \textsc{Fix} children are added with compiler error messages as feedback (Line 13). This feedback-driven error repair enables LAC2R to progressively converge to correct translations (Lines 4--13).

\begin{algorithm}[t]
\caption{High-level procedure of LAC2R's MCTS}
\label{alg:mcts-translation}
\setcounter{AlgoLine}{-1}
\SetKwFunction{MCTSSearch}{MCTS\_Search}
\SetKwFunction{MCTSRollout}{MCTS\_Rollout}
\SetKwFunction{FindBestSolution}{Find\_Best\_Solution}
\SetKwFunction{Select}{Select}
\SetKwFunction{Expand}{Expand}
\SetKwFunction{Simulate}{Simulate}
\SetKwFunction{Backpropagate}{Backpropagate}
\KwIn{$func$: target Rust function; $program$: full Rust program;\\
      \phantom{\textbf{Input: }}$N_{rollout}$: number of MCTS rollouts;\\
      \phantom{\textbf{Input: }}$S(\cdot)$: safety ratio function (fraction of safe code in a Rust snippet);\\
      \phantom{\textbf{Input: }}$Q(n)$: accumulated reward of node $n$}
\KwOut{$best\_node$: highest-reward solution node found}
\textbf{Node structure:} each node $n$ holds
  $n.rust$ (translated Rust snippet),
  $n.program$ (full Rust program with $n.rust$ substituted),
  $n.type \in \{\textsc{Init}, \textsc{Gen}, \textsc{Fix}, \textsc{Success}\}$\;
\SetKwProg{Pn}{Func}{:}{}
\Pn{\MCTSSearch{$func$, $program$}}{
    $root \gets$ \texttt{Create\_Node}($func$, type$=$\textsc{Init})\tcp*{root node; no LLM call yet}
    \For{$i \gets 1$ \textbf{to} $N_{rollout}$}{
        \MCTSRollout{$root$, $i$, $program$}\;
    }
    \KwRet \FindBestSolution{$root$}\;
}
\Pn{\MCTSRollout{$root$, $i$, $program$}}{
    $path_s \gets$ \Select{$root$, $i$}\tcp*{UCT-guided path to a leaf}
    $leaf \gets path_s.\mathrm{last}()$\;
    \Expand{$leaf$, $program$}\tcp*{generate children; see Alg.~\ref{alg:expand}}
    $path_r \gets$ \Simulate{$leaf$}\tcp*{greedy rollout from leaf to terminal}
    \Backpropagate{$path_s \mathbin{+\!\!+} path_r$}\tcp*{update $Q$ and visit counts along full path}
    \lIf{$path_r \neq \emptyset$}{\KwRet $path_r.\mathrm{last}()$}
    \lElse{\KwRet $leaf$}
}
\Pn{\FindBestSolution{$root$}}{
    $\mathcal{S} \gets \{n \in \text{tree}(root) \mid n.type = \textsc{Success}\}$\tcp*{all test-passing nodes}
    \eIf{$\mathcal{S} \neq \emptyset$}{
        \KwRet $\arg\max_{n \in \mathcal{S}}\; S(n.rust)$\tcp*{highest safety ratio among valid solutions}
    }{
        \KwRet $root$\tcp*{no valid solution found; return original input}
    }
}
\end{algorithm}

\begin{algorithm}[h]
\caption{High-level procedure of \texttt{Expand}()}\label{alg:expand}

\SetKwFunction{FExpand}{Expand}
\SetKwFunction{AddChild}{Add\_Child}

\setcounter{AlgoLine}{-1}
\KwIn{$node$: MCTS node; $program$: full Rust program; $T^{test}$: test suite (may be $\emptyset$)}
\nl \textbf{Types:} \textsc{Init} (root), \textsc{Gen} (initial translation), \textsc{Fix} (error repair), \textsc{Success} (valid solution)\;

\SetKwProg{Pn}{Func}{:}{}

\Pn{\FExpand{$node$, $program$}}{
    \eIf{$node.type = \textsc{Init}$\tcp*[f]{root: spawn diverse initial candidates}}{
        create multiple \textsc{Gen} children of $node$ using heterogeneous LLMs\;
    }{
        $resp \gets \texttt{LLM}(node.conversation)$\tcp*{translate using conversation history and prior feedback}
        insert $resp$ into $program$\;
        \eIf{\texttt{compile}($program$) succeeds}{
            $node.reward \gets \texttt{compute\_reward}(program)$\tcp*{based on safety score and compile status}
            \eIf{$|T^{test}|>0$}{
                \eIf{\texttt{run\_tests}($program$, $T^{test}$) passes}{
                    add \textsc{Success} child to $node$\tcp*{solution candidate found}
                }{
                    add \textsc{Fix} children to $node$ with test-run errors as feedback\;
                }
            }{
                add \textsc{Success} child to $node$\tcp*{no tests: compile pass is sufficient}
            }
        }{
            add \textsc{Fix} children to $node$ with compile errors as feedback\;
        }
    }
}
\end{algorithm}

\begin{figure}[hbpt]
    \centering  
    \includegraphics[width=1.0\linewidth]{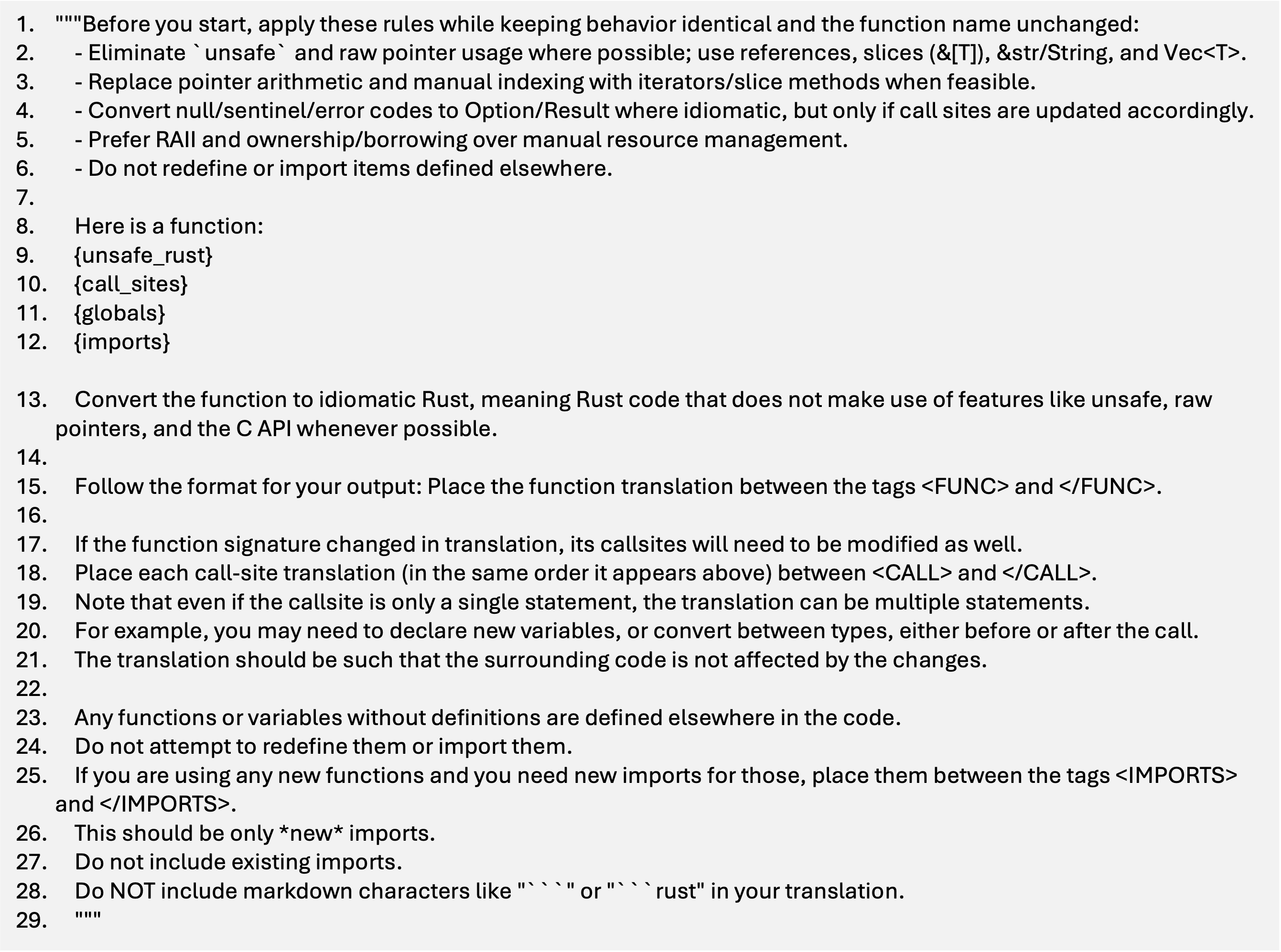} 
    \Description{Decorative image; no additional information.}
    \caption{Main prompt used in LAC2R. When compilation or test-run errors occur, the error messages are appended to the conversation history and passed back to the LLM as feedback for the next \textsc{Fix}-node expansion.}
    \label{fig:main_prompt}
\end{figure}

\noindent \textbf{Prompt.} Figure~\ref{fig:main_prompt} presents the main prompt for LAC2R. The prompt largely follows a prior work, \textsc{C2SaferRust}'s prompt except for line 1-6 that explicitly clarifies the translation direction. The prompt defines four placeholders in line 9-12. \texttt{\{unsafe\_rust\}} is for the target Rust code, \texttt{\{call\_sites\}} is for the list of call sites calling the target Rust function, \texttt{\{global\}} is for global variables, and \texttt{\{import\}} is for import information. The prompt instructs LLMs to produce the translation results within \texttt{<FUNC>} and \texttt{<$\backslash$FUNC>}. In addition to the main prompt, LAC2R employs two more prompts, one for correcting compile errors and another for addressing testcase execution errors. Both prompts follow the \textsc{C2SaferRust} prompts.

\section{Experiments}
\label{sec:experiment}

\subsection{Experimental Setup}

\textbf{Baseline and Benchmarks.} For experimental comparison, we selected \textsc{C2SaferRust} as our baseline, as its open implementation has been empirically shown to support large-scale, real-world benchmarks, as shown in Table~\ref{tbl:llm-approaches-experiment}. \textsc{C2SaferRust} supports two types of decompositions: function-wise and unit-wise, where a unit can be smaller than a function. In their experiments, the unit-wise translation tended to outperform the other. We conduct our experiments using both variants of \textsc{C2SaferRust}. Notably, LAC2R adopts function-wise decomposition and can be easily extended to support the unit-wise translation. In addition to \textsc{C2SaferRust}, we selected \textsc{EvoC2Rust} as another counterpart. Unlike both \textsc{C2SaferRust} and LAC2R, \textsc{EvoC2Rust} adopts a skeleton-guided translation strategy and does not rely on conventional transpiler-based preprocessing.

For comparison on large-scale, repository-level benchmarks, we conducted experiments on two datasets: seven C programs collected from GNU coreutils and ten C programs used by a prior transpiler, Laertes~\cite{10.1145/3485498}, that \textsc{C2SaferRust} introduced. The maximum LoCs in these two benchmark datasets are 14K and 96K, respectively. For comparison on small-scale benchmarks, we conducted experiments on the \textit{TRACTOR Public-Tests}~\cite{darpa_tractor_corpus} benchmark, which comprises Battery-01 (\texttt{B01\_organic}: 38 programs and \texttt{B01\_synthetic}: 85 programs) and one project-level program (\texttt{P00\_perlin\_noise}), for a total of 124 programs. TRACTOR Public-Tests was planned, curated, and created by MIT Lincoln Laboratory, the independent test and evaluation (T\&E) organization for the DARPA TRACTOR program, and was designed to objectively evaluate the artifacts produced by participating performer teams. Battery-01 in total contains 150 benchmarks, of which 123 are public and 27 are reserved as hidden test cases for independent evaluation. \texttt{P00\_perlin\_noise} is an independent project-level benchmark outside the Battery-01 suite, included in this work to broaden the evaluation to project-scale translation.

\begin{table*}
  \caption{Prior LLM-based approaches for C-to-Rust translation with their benchmarks and primary metrics. }
  \label{tbl:llm-approaches-experiment}
  \small
  \centering
  \begin{tabular}{lll}
    \toprule
    \textit{Name}     & \textit{Benchmarks (LoC)}     & \makecell[l]{\textit{Primary metrics}} \\
    \midrule
    \midrule
    \makecell[l]{VERT~\cite{yang2024vertverifiedequivalentrust}} & \makecell[l]{TransCoder-IR dataset(<100), 14 PM prog.\textsuperscript{1} (<500)}  & \makecell[l]{success rates, runtime}  \\
    \midrule
    \makecell[l]{F{\small LOURINE}~\cite{eniser2024translatingrealworldcodellms} }     & \makecell[l]{libopenaptx (<200), opl (<500)} & \makecell[l]{success rates, linter warnings}  \\
    \midrule
    \makecell[l]{S{\small PEC}T{\small RA}~\cite{nitin2024spectraenhancingcodetranslation}}   & CodeNet dataset    & success rates \\
    \midrule
    \makecell[l]{S{\small YZYGY}~\cite{shetty2024syzygydualcodetestc} }     & \makecell[l]{Zopfli (<3K), urlparser (<500)}       & \makecell[l]{success rates, slow down} \\
    \midrule
    \makecell[l]{\textsc{C2SaferRust}~\cite{nitin2025c2saferrusttransformingcprojects}}    & \makecell[l]{coreutils dataset (<15K), Laertes dataset (<96K)}       & \makecell[l]{unsafe constructs\textsuperscript{2}, success rates} \\
    \midrule
    \makecell[l]{SACTOR~\cite{zhou2025llmdrivenmultisteptranslationc} }     & \makecell[l]{TransCoder-IR dataset(<100), CodeNet dataset(<500), \\avl-tree (<500), urlparser (<500)}       & \makecell[l]{success rates, cost in tokens, linter warnings } \\
    \midrule
    \makecell[l]{ SafeTrans~\cite{farrukh2025safetransllmassistedtranspilationc} }  & \makecell[l]{ 2,653 C programs of CodeNet dataset (<500)}       & \makecell[l]{ success rates, coverages, memory safety } \\
    \midrule
    \makecell[l]{ LLMigrate~\cite{liu2025llmigratetransforminglazylarge} }   & \makecell[l]{ Three Linux kernel modules: math, sort, ramfs (<900)}       & \makecell[l]{success rates, safe code ratio, laziness, C-BLEU } \\    \midrule
    \makecell[l]{ EvoC2Rust~\cite{wang2025evoc2rustskeletonguidedframeworkprojectlevel} }   & \makecell[l]{ Vivo-Bench (<1K), C2R-Bench (<4K)}       & \makecell[l]{ success rates, translation accuracy, memory-safe rate} \\    \midrule
    \makecell[l]{ \textsc{RustAssure}~\cite{11334436} }   & \makecell[l]{ libcsv(<500), urlparser(<500), optipng(<3.1K), libbmp(<300), u8c(<400)}       & \makecell[l]{compile rate, unsafe code rate, semantic similarity score} \\    \midrule    
    \makecell[l]{ \textsc{Rustify}~\cite{wang2026rustify} }   & \makecell[l]{ HumanEval dataset(<500), urlparser, libtree (<2K), json.h(<3.5K), \\ quadtree(<500), buffer(<500), rgba(<500), binn(<4.5K), private (<10K)}       & \makecell[l]{pass rate, compile rate, unsafe ratio, codeBLEU} \\    \midrule
    \makecell[l]{\textsc{RustMap}~\cite{Cai25:Rustmap}}    & \makecell[l]{RosettaCode dataset, bzip2 (<45K) }       & \makecell[l]{coverage, unsafe code blocks, cognitive complexity} \\    \bottomrule
    \multicolumn{3}{l}{1) PM stands for Pointer Manipulation programs. }\\
    \multicolumn{3}{l}{2) The unsafe constructs include raw pointer declarations, raw pointer dereferences, unsafe lines, unsafe type casts, and unsafe calls in Rust codes.}\\
\end{tabular}
\end{table*}

\textbf{Metrics.} We primarily measure the counts of five unsafe Rust constructs, such as $RPD_1$, $RPD_2$, $LUC$, $UCE$, and $UTC$, as introduced in~\cite{nitin2025c2saferrusttransformingcprojects}. We then compute the \textit{Safety Ratio} ($SR$) as a function of these five counts using Equation~\ref{eq:safety}. We also compute the \textit{Function-level Compile Rate} ($FCR$), defined as the number of functions that compile successfully after translation divided by the total number of functions in a given benchmark. We next measure the \textit{Function Replacement Rate} ($FRR$), defined as the proportion of functions successfully translated in a given benchmark, where a translation is considered successful when there are no compile-time or testcase execution errors. We also measure the \textit{Test Pass Rate} ($TPR$), defined as the number of passed test cases divided by the total number of test cases. For additional code quality evaluation, we measure the number of \textit{Linter Warnings} in the final translation using \texttt{Clippy}~\cite{rustclippy}, which checks the idiomaticity and quality of Rust code. \texttt{Clippy} supports multiple lint levels to help catch various mistakes, so that the number of linter warnings is not necessarily proportional to the number of unsafe Rust constructs. In addition, to assess the translation costs, we measure the average number of tokens consumed by LLMs and the number of LLM queries for translating a function of a given benchmark. 

For small-scale code datasets, we measure \textit{Program-level Compile Rate} ($PCR$), defined as the number of programs that compile successfully after translation divided by the total number of programs in a given dataset. In addition, We also measure the \textit{Program Pass Rate} ($PPR$), defined as the number of programs that pass all tests after translation divided by the total number of programs.

\textbf{LLMs.} For comparison, counterpart approaches use GPT-4o, whereas LAC2R, leverages the pair of GPT-4o and DeepSeek V3~\cite{deepseekv3} for LAC2R. LLM capability is often correlated to its scale, although architecture and training also matter. The estimated numbers of parameters for GPT-4o and DeepSeek V3 are 200B and 671B~\cite{abacha2025medecbenchmarkmedicalerror, deepseekv3}.

\textbf{MCTS Configuration.} LAC2R runs $N_{rollout} = 10$ rollouts per function with a UCT exploration constant of $C = 1.5$ (Algorithm~\ref{alg:mcts-translation}). The maximum tree depth is set to $D = 5$. The root Init node generates four Gen children (two from GPT-4o and two from DeepSeek V3) as in Algorithm~\ref{alg:expand}. The reward weight is set to $w = 2$, giving twice the emphasis to safety improvement over compile success.

\textbf{Machine and Running Time for Experiments.} We ran all experiments on a workstation equipped with Intel Xeon Gold 6434 (32 cores), 512GB memory, and Ubuntu 22.04.4 LTS. These specifications are reported for reproducibility, not as a requirement. CPU/memory utilization remained modest in our experiments. The longest benchmark of coreutils is \texttt{tail}, for which LAC2R completed its translation in approximately 14 hours. In the Laertes dataset, the longest benchmark is \texttt{optipng}. LAC2R completed its translation in 48 hours. Between the two variants of \textsc{C2SaferRust}, the one with the unit-wise decomposition runs longer time than the other. \textsc{C2SaferRust} with 150 LoC-unit size translated \texttt{optipng} in approximately 120 hours. Elapsed time was largely driven by refinement iterations and external LLM latency. \textsc{EvoC2Rust} translated \texttt{tail} in approximately 1.4 hours and \texttt{optipng} in approximately 2.7 hours.

\subsection{Results}

\begin{figure}[hbpt]
    \centering  
    \includegraphics[width=1.0\linewidth]{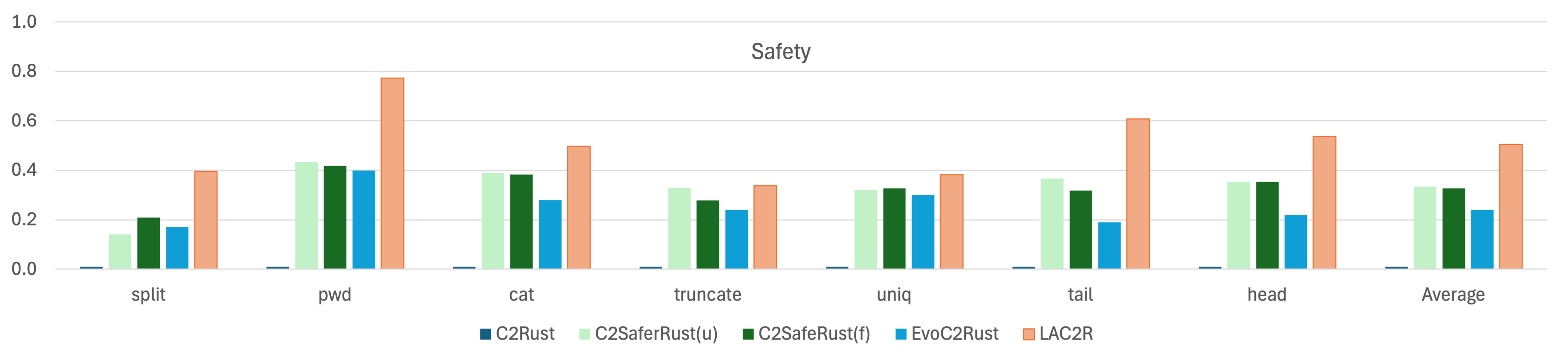} 
    \Description{Decorative image; no additional information.}
    \caption{Safety Ratio: visualization of the $SR$ column in Table~\ref{tab:main_exp_coreutils}  for GNU coreutils benchmarks. }
    \label{fig:main_exp_coreutils_safety}
\end{figure}

\begin{figure}[hbpt]
    \centering  
    \includegraphics[width=1.0\linewidth]{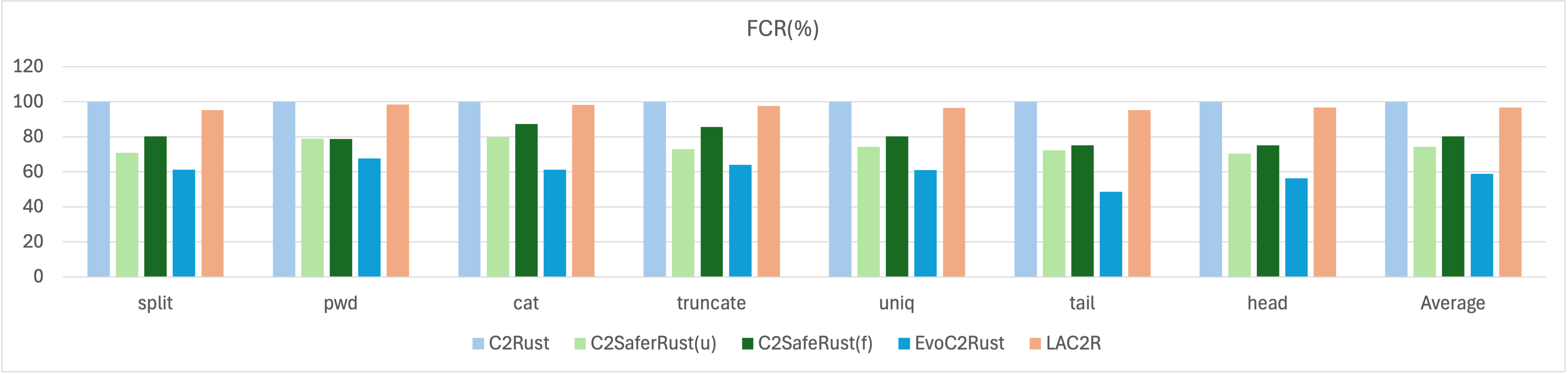} 
    \Description{Decorative image; no additional information.}
    \caption{Function Compile Rate: visualization of the $FCR$ column in Table~\ref{tab:main_exp_coreutils} for GNU coreutils benchmarks. }
    \label{fig:main_exp_coreutils_sr}
\end{figure}

\begin{figure}[hbpt]
    \centering  
    \includegraphics[width=1.0\linewidth]{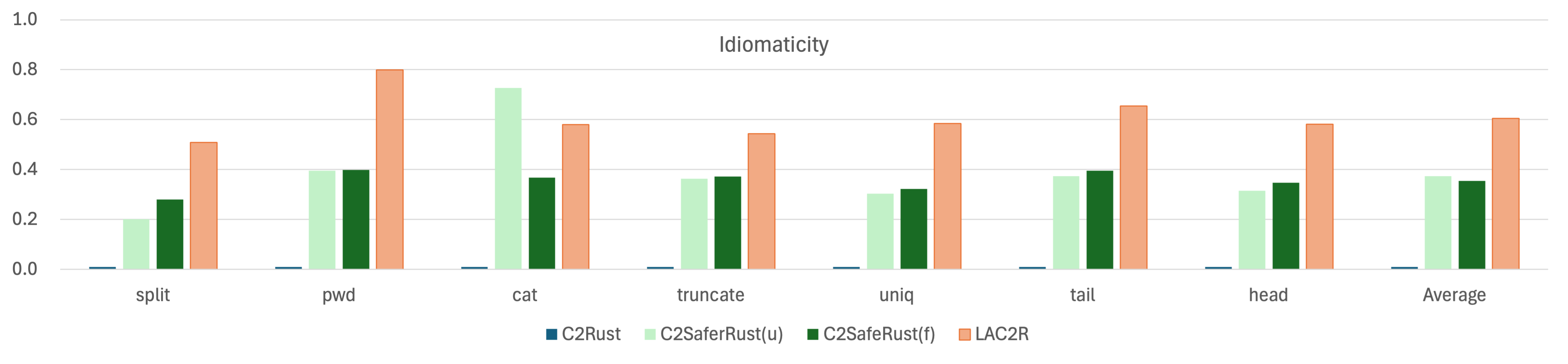} 
    \Description{Decorative image; no additional information.}
    \caption{Idiomaticity: visualization of $I$ computed using the Linter warnings in Table~\ref{tab:main_exp_coreutils} for GNU coreutils benchmarks. }
    \label{fig:main_exp_coreutils_safety_linter}
\end{figure}

\textbf{Comparative Evaluation on large-scale Benchmarks.} We evaluated LAC2R against the counterparts on two large-scale benchmarks, such as the coreutils and Laertes. Table~\ref{tab:main_exp_coreutils} shows the results on the coreutils dataset in several metrics, such as $SR$, $FCR$, $FRR$, $TPR$, LLM costs, and linter warnings. On most coreutil benchmarks, LAC2R outperforms the counterparts by a large margin. Figure~\ref{fig:main_exp_coreutils_safety} and~\ref{fig:main_exp_coreutils_sr} visualize the results. To normalize Linter warnings, we define a linter-based idiomaticity score $I$ as: 
\begin{align}
\label{eq:safety_linter}
&I(r_{i})= max\bigg\{ \bigg( 1-\frac{N_{L}(r_i)}{N_{L}(r_0)}  \bigg),0 \bigg\},
\end{align}
where $N_{L}(r_i)$ denotes the number of linter warnings for Rust code $r_i$. The comparison in $I$ is presented in Figure~\ref{fig:main_exp_coreutils_safety_linter}, that shows that LAC2R generally scores higher and LAC2R produces Rust code with better idiomaticity compared to its counterparts.

As in Table~\ref{tab:main_exp_coreutils}, \textsc{LAC2R} achieves an average $SR$ of 0.50 across coreutil benchmarks, compared to 0.33 for \textsc{C2SaferRust}(U), 0.32 for \textsc{C2SaferRust}(F), and 0.24 for \textsc{EvoC2Rust}, a gain of up to 52\% over the strongest safety baseline. This consistent advantage reflects LAC2R's safety-aware reward design (Algorithm~\ref{alg:expand}): by weighting safety improvement $R_S$ twice as heavily as compile success $R_V$, UCT selection (Algorithm~\ref{alg:mcts-translation}) preferentially explores refinement paths that reduce unsafe constructs rather than merely achieving compilation. Translations that compile but retain raw-pointer operations receive lower rewards, steering the search toward safer code across ten rollouts.

Table~\ref{tab:main_exp_coreutils} also shows the translation cost, including the number of LLM queries and tokens consumed. LAC2R incurs more queries and higher token consumption than its counterpart, which is expected given that LAC2R leverages the MCTS framework to search for better decision trajectories over a candidate population generated by LLMs.

\begin{table}[b]
\caption{Comparative results on coreutils in detail. \textsc{C2SaferRust}(U) and \textsc{C2SaferRust}(F) denote two versions of \textsc{C2SaferRust}: the former uses unit-wise decomposition, and the latter employs function-wise decomposition. Bold highlights the best performance across the approaches.}
\centering
\label{tab:main_exp_coreutils}
\begin{tiny}
\begin{tabular}{|c|c|c|c|c|c|c|c|c|}
\hline
\begin{tabular}[c]{@{}c@{}}\textit{Benchmark}\\(LoC)\end{tabular} &
\textit{Methods} &
SR &
\begin{tabular}[c]{@{}c@{}}FCR\end{tabular} &
\begin{tabular}[c]{@{}c@{}}FRR\end{tabular} &
\begin{tabular}[c]{@{}c@{}}TPR\end{tabular} &
\begin{tabular}[c]{@{}c@{}}Linter\\Warns\end{tabular} &
\begin{tabular}[c]{@{}c@{}}Ave.\\Query\end{tabular} &
\begin{tabular}[c]{@{}c@{}}Ave.\\Tokens\end{tabular} \\ \hline\hline

\multirow{6}{*}{\begin{tabular}[c]{@{}c@{}}split\\(13,848)\end{tabular}}
& \textsc{C2Rust} & 0 & - & - & 100\% & 546 & - & - \\ \cline{2-9}
& \textsc{C2SaferRust(u)} & 0.14 & 70.89\% & 38.70\% & 100\% & 436 & 3.12 & 5,414 \\ \cline{2-9}
& \textsc{C2SaferRust(f)} & 0.20 & 80.19\% & 71.98\% & 100\% & 393 & \textbf{2.32} & \textbf{2,961} \\ \cline{2-9}
& \textsc{EvoC2Rust} & 0.17 & 61.10\% & 61.10\% & 0\% & - & 1.01 & 3,125 \\ \cline{2-9}
& LAC2R & \textbf{0.39} & \textbf{95.16\%} & \textbf{89.37}\% & 100\% & \textbf{268} & 18.82 & 40,536 \\ \hline\hline

\multirow{6}{*}{\begin{tabular}[c]{@{}c@{}}pwd\\(5,859)\end{tabular}}
& \textsc{C2Rust} & 0 & - & - & 100\% & 369 & - & - \\ \cline{2-9}
& \textsc{C2SaferRust(u)} & 0.43 & 78.95\% & 77.63\% & 100\% & 223 & \textbf{2.55} & \textbf{3,842} \\ \cline{2-9}
& \textsc{C2SaferRust(f)} & 0.41 & 78.74\% & 75.59\% & 100\% & 222 & 2.59 & 4,045 \\ \cline{2-9}
& \textsc{EvoC2Rust} & 0.40 & 67.60\% & 67.60\% & 0\%  & - & 1.00 & 3,068 \\ \cline{2-9}
& LAC2R & \textbf{0.77} & \textbf{98.43\%} & \textbf{97.63}\% & 100\% & \textbf{74} & 17.10 & 50,170 \\ \hline\hline

\multirow{6}{*}{\begin{tabular}[c]{@{}c@{}}cat\\(7,460)\end{tabular}}
& \textsc{C2Rust} & 0 & - & - & 100\% & 414 & - & - \\ \cline{2-9}
& \textsc{C2SaferRust(u)} & 0.39 & 79.51\% & 76.59\% & 100\% & \textbf{113} & \textbf{2.37} & 3,763 \\ \cline{2-9}
& \textsc{C2SaferRust(f)} & 0.38 & 87.35\% & 83.73\% & 100\% & 262 & 2.42 & \textbf{3,554} \\ \cline{2-9}
& \textsc{EvoC2Rust} & 0.28 & 61.2\% & 61.2\% &  0\% & - & 1.51 & 4,182 \\ \cline{2-9}
& LAC2R & \textbf{0.49} & \textbf{98.19\%} & \textbf{95.18}\% & 100\% & 174 & 14.26 & 28,851 \\ \hline\hline

\multirow{6}{*}{\begin{tabular}[c]{@{}c@{}}truncate\\(7,181)\end{tabular}}
& \textsc{C2Rust} & 0 & - & - & 100\% & 385 & - & - \\ \cline{2-9}
& \textsc{C2SaferRust(u)} & 0.32 & 72.94\% & 68.24\% & 100\% & 245 & \textbf{2.63} & 4,595 \\ \cline{2-9}
& \textsc{C2SaferRust(f)} & 0.27 & 85.48\% & 80.65\% & 100\% & 242 & 2.69 & \textbf{4,264} \\ \cline{2-9}
& \textsc{EvoC2Rust} & 0.24 & 63.9\% & 63.9\% & 0\% & - & 1.02 & 3,036 \\ \cline{2-9}
& LAC2R & \textbf{0.33} & \textbf{97.58\%} & \textbf{92.74}\% & 100\% & \textbf{176} & 15.25 & 36,963 \\ \hline\hline

\multirow{6}{*}{\begin{tabular}[c]{@{}c@{}}uniq\\(8,299)\end{tabular}}
& \textsc{C2Rust} & 0 & - & - & 100\% & 469 & - & - \\ \cline{2-9}
& \textsc{C2SaferRust(u)} & 0.32 & 74.18\% & 71.36\% & 100\% & 327 & 2.59 & 4,738 \\ \cline{2-9}
& \textsc{C2SaferRust(f)} & 0.32 & 80.24\% & 76.65\% & 100\% & 318 & \textbf{2.48} & \textbf{3,541} \\ \cline{2-9}
& \textsc{EvoC2Rust} & 0.30 & 61.00\% & 61.00\% & 0\% & - & 1.01 & 3,103 \\ \cline{2-9}
& LAC2R & \textbf{0.38} & \textbf{96.41\%} & \textbf{94.61}\% & 100\% & \textbf{195} & 15.91 & 30,742 \\ \hline\hline

\multirow{6}{*}{\begin{tabular}[c]{@{}c@{}}tail\\(14,423)\end{tabular}}
& \textsc{C2Rust} & 0 & - & - & 100\% & 717 & - & - \\ \cline{2-9}
& \textsc{C2SaferRust(u)} & 0.36 & 72.29\% & 69.43\% & 100\% & 449 & \textbf{2.50} & 4,707 \\ \cline{2-9}
& \textsc{C2SaferRust(f)} & 0.31 & 75.19\% & 71.80\% & 100\% & 434 & 2.51 & \textbf{3,768} \\ \cline{2-9}
& \textsc{EvoC2Rust} & 0.19 & 48.5\% & 48.5\% & 0\% & - & 1.02 & 3,170 \\ \cline{2-9}
& LAC2R & \textbf{0.60} & \textbf{95.11\%} & \textbf{93.60}\% & 100\% & \textbf{248} & 16.65 & 42,187 \\ \hline\hline

\multirow{6}{*}{\begin{tabular}[c]{@{}c@{}}head\\(8,047)\end{tabular}}
& \textsc{C2Rust} & 0 & - & - & 100\% & 409 & - & - \\ \cline{2-9}
& \textsc{C2SaferRust(u)} & 0.35 & 70.47\% & 66.84\% & 100\% & 280 & 2.67 & 4,523 \\ \cline{2-9}
& \textsc{C2SaferRust(f)} & 0.35 & 75.16\% & 70.59\% & 100\% & 267 & \textbf{2.60} & \textbf{4,220} \\ \cline{2-9}
& \textsc{EvoC2Rust} & 0.22 & 56.2\% & 56.2\% & 0\% & - & 1.02 & 4,234 \\ \cline{2-9}
& LAC2R & \textbf{0.53} & \textbf{96.73\%} & \textbf{92.15}\% & 100\% & \textbf{171} & 15.94 & 35,988 \\ \hline\hline

\textbf{Average}
& \textsc{C2Rust} & 0 & - & - & 100\% & 472 & - & - \\ \cline{2-9}
& \textsc{C2SaferRust(u)} & 0.33 & 74.17\% & 66.97\% & 100\% & 296 & 2.63 & 4,512 \\ \cline{2-9}
& \textsc{C2SaferRust(f)} & 0.32 & 80.33\% & 75.85\% & 100\% & 305 & 2.51 & 3,765 \\ \cline{2-9}
& \textsc{EvoC2Rust} & 0.24 & 58.90\% & 58.90\% & 0\% & - & \textbf{1.02} & \textbf{3,220} \\ \cline{2-9}
& LAC2R & \textbf{0.50} & \textbf{96.80\%} & \textbf{93.61}\% & 100\% & \textbf{186} & 16.27 & 37,920 \\ \hline

\end{tabular}
\end{tiny}
\end{table}

On most Laertes benchmarks, LAC2R outperforms the counterparts in the metrics including $SR$, $FCR$, $FRR$, $TPR$ and linter warnings as in Table~\ref{tab:main_exp_laertes}. These results indicate that LAC2R more effectively reduces the unsafe Rust constructs than its counterpart during translation without sacrificing $FRR$ or $TPR$. Table~\ref{tab:main_exp_laertes} also shows that LAC2R incurs more queries and higher token consumption than its counterpart due to its MCTS-based structure. 

For \texttt{bzip2} in Table~\ref{tab:main_exp_laertes}, LAC2R produces more linter warnings than C2Rust, including an additional 234 instances of the "deref which would be done by auto-deref" warning. This reflects cases where the LLM does not leverage Rust's automatic dereferencing mechanism when translating C-style unsafe code to safe Rust, and instead emits explicit dereference operations in otherwise safe code. These warnings are not part of the five categories of unsafe constructs. Rather, they occur in safe but non-idiomatic Rust. Moreover, \textsc{C2SaferRust}(U) yields more linter warnings than C2Rust, driven by 417 instances of the "unnecessary \texttt{unsafe} block" warning, indicating the insertion of redundant \texttt{unsafe} blocks.

\begin{table}[htbp]
\caption{Comparative results on Laertes benchmarks in detail. \textsc{C2SaferRust}(U) and \textsc{C2SaferRust}(F) denote two versions of \textsc{C2SaferRust}: the former uses unit-wise decomposition, and the latter employs function-wise decomposition. Bold highlights the best performance across the approaches.}
\centering
\label{tab:main_exp_laertes}
\begin{tiny}
\begin{tabular}{|c|c|c|c|c|c|c|c|}
\hline
\begin{tabular}[c]{@{}c@{}} \textit{Benchmark} \\ (LoC) \end{tabular} &
\multicolumn{1}{c|}{\textit{Methods}} &
\begin{tabular}[c]{@{}c@{}}SR\end{tabular} &
\begin{tabular}[c]{@{}c@{}} FCR \end{tabular} &
\begin{tabular}[c]{@{}c@{}} Project \\ Compilability \end{tabular} &
\begin{tabular}[c]{@{}c@{}} Linter \\ Warns \end{tabular} &
\begin{tabular}[c]{@{}c@{}} Ave. \\ Query \end{tabular} &
\begin{tabular}[c]{@{}c@{}} Ave. \\ Tokens \end{tabular} \\ \hline \hline

\multirow{5}{*}{\begin{tabular}[c]{@{}c@{}} bzip2 \\ (43,374) \end{tabular}}
    & \textsc{C2Rust}                 & 0            & -            & \checkmark   & 458          & -              & -        \\ \cline{2-8}
    & \textsc{C2SaferRust(u)}         & 0.23         & 61.75\%          & \checkmark   & 813          & 2.75           & 9,749    \\ \cline{2-8}
    & \textsc{C2SaferRust(f)}         & 0.35         & 62.50\%          & \checkmark   & \textbf{442} & 2.91           & 11,452   \\ \cline{2-8}
    & \textsc{EvoC2Rust}              & 0.33         & 79.50\%          & $\times$     & -            & \textbf{1.01}  & \textbf{3,271}  \\ \cline{2-8}
    & LAC2R                           & \textbf{0.70} & \textbf{85.16}\% & \checkmark   & 544          & 18.99          & 85,380   \\ \hline \hline

\multirow{5}{*}{\begin{tabular}[c]{@{}c@{}} genann \\ (2,084) \end{tabular}}
    & \textsc{C2Rust}                 & 0            & -            & \checkmark   & 114          & -              & -        \\ \cline{2-8}
    & \textsc{C2SaferRust(u)}         & 0.31         & 62.86\%          & \checkmark   & 129          & 2.27           & 5,473    \\ \cline{2-8}
    & \textsc{C2SaferRust(f)}         & 0.31         & 59.38\%          & \checkmark   & 142          & 2.89           & 13,185   \\ \cline{2-8}
    & \textsc{EvoC2Rust}              & \textbf{0.72} & 63.00\%         & $\times$     & -            & \textbf{1.02}  & \textbf{4,482}  \\ \cline{2-8}
    & LAC2R                           & 0.61         & \textbf{93.75}\% & \checkmark   & \textbf{113} & 19.33          & 105,465  \\ \hline \hline

\multirow{5}{*}{\begin{tabular}[c]{@{}c@{}} lil \\ (5,400) \end{tabular}}
    & \textsc{C2rust}                 & 0            & -            & \checkmark   & 419          & -              & -        \\ \cline{2-8}
    & \textsc{C2SaferRust(u)}         & 0.06         & 27.50\%          & \checkmark   & 390          & 2.93           & 5,976    \\ \cline{2-8}
    & \textsc{C2SaferRust(f)}         & 0.20         & 35.00\%          & \checkmark   & 442          & 3.25           & 8,067    \\ \cline{2-8}
    & \textsc{EvoC2Rust}              & 0.06         & 27.10\%          & $\times$     & -            & \textbf{1.03}  & \textbf{4,899}  \\ \cline{2-8}
    & LAC2R                           & \textbf{0.30} & \textbf{74.38}\% & \checkmark   & \textbf{221} & 22.09          & 78,088   \\ \hline \hline

\multirow{5}{*}{\begin{tabular}[c]{@{}c@{}} urlparser \\ (1,118) \end{tabular}}
    & \textsc{C2Rust}                 & 0            & -            & \checkmark   & 94           & -              & -        \\ \cline{2-8}
    & \textsc{C2SaferRust(u)}         & 0.20         & 52.00\%          & \checkmark   & 93           & 3.08           & 8,400    \\ \cline{2-8}
    & \textsc{C2SaferRust(f)}         & 0.34         & 50.00\%          & \checkmark   & 64           & 3.45           & 13,772   \\ \cline{2-8}
    & \textsc{EvoC2Rust}              & \textbf{0.83} & \textbf{85.7\%}  & $\times$     & -            & \textbf{1.00}  & \textbf{4,267}  \\ \cline{2-8}
    & LAC2R                           & 0.50         & 81.81\%          & \checkmark   & \textbf{63}  & 17.33          & 122,033  \\ \hline \hline

\multirow{5}{*}{\begin{tabular}[c]{@{}c@{}} grabc \\ (1,046) \end{tabular}}
    & \textsc{C2Rust}                 & 0            & -            & \checkmark   & 45           & -              & -        \\ \cline{2-8}
    & \textsc{C2SaferRust(u)}         & 0.19         & 71.43\%          & \checkmark   & 35           & 2.40           & 8,272    \\ \cline{2-8}
    & \textsc{C2SaferRust(f)}         & 0.20         & 71.43\%          & \checkmark   & 36           & 2.60           & \textbf{3,744}  \\ \cline{2-8}
    & \textsc{EvoC2Rust}              & \textbf{0.70} & 45.8\%          & $\times$     & -            & \textbf{1.00}  & 3,854    \\ \cline{2-8}
    & LAC2R                           & 0.61         & \textbf{100}\%   & \checkmark   & \textbf{24}  & 22.00          & 56,913   \\ \hline \hline

\multirow{5}{*}{\begin{tabular}[c]{@{}c@{}} tulip-\\indicators \\ (44,486) \end{tabular}}
    & \textsc{C2Rust}                 & 0            & -            & \checkmark   & 806          & -              & -        \\ \cline{2-8}
    & \textsc{C2SaferRust(u)}         & 0.13         & 81.82\%          & \checkmark   & \textbf{574} & 2.75           & 13,239   \\ \cline{2-8}
    & \textsc{C2SaferRust(f)}         & 0.11         & 81.20\%          & \checkmark   & 608          & 2.38           & 7,263    \\ \cline{2-8}
    & \textsc{EvoC2Rust}              & \textbf{0.54} & 71.9\%          & $\times$     & -            & \textbf{1.06}  & \textbf{4,367}  \\ \cline{2-8}
    & LAC2R                           & 0.23         & \textbf{91.03}\% & \checkmark   & 591          & 15.41          & 60,995   \\ \hline \hline

\multirow{5}{*}{\begin{tabular}[c]{@{}c@{}} optipng \\ (95,560) \end{tabular}}
    & \textsc{C2Rust}                 & 0            & -            & \checkmark   & 5270         & -              & -        \\ \cline{2-8}
    & \textsc{C2SaferRust(u)}         & 0.45         & 50.41\%          & \checkmark   & 7,184        & 2.34           & \textbf{4,355}  \\ \cline{2-8}
    & \textsc{C2SaferRust(f)}         & 0.11         & 48.78\%          & \checkmark   & 5164         & 2.75           & 10,361   \\ \cline{2-8}
    & \textsc{EvoC2Rust}              & \textbf{0.90} & 64.8\%          & $\times$     & -            & \textbf{1.73}  & 5,733    \\ \cline{2-8}
    & LAC2R                           & 0.33         & \textbf{89.57}\% & \checkmark   & \textbf{3,609} & 21.47        & 118,386  \\ \hline \hline

\multirow{5}{*}{\begin{tabular}[c]{@{}c@{}} qsort \\ (41) \end{tabular}}
    & \textsc{C2Rust}                 & 0            & -           & \checkmark   & 5            & -              & -        \\ \cline{2-8}
    & \textsc{C2SaferRust(u)}         & 0.72         & 66.67\%          & \checkmark   & 4            & 2.50           & 3,442    \\ \cline{2-8}
    & \textsc{C2SaferRust(f)}         & \textbf{1}   & \textbf{100}\%   & \checkmark   & 1            & 1.67           & \textbf{1,552}  \\ \cline{2-8}
    & \textsc{EvoC2Rust}              & \textbf{1.00} & \textbf{100}\%  & $\times$     & -            & \textbf{1.00}  & 5,107    \\ \cline{2-8}
    & LAC2R                           & \textbf{1}   & \textbf{100}\%   & \checkmark   & \textbf{0}   & 15.00          & 22,534   \\ \hline \hline

\multirow{5}{*}{\begin{tabular}[c]{@{}c@{}} snudown \\ (6,521) \end{tabular}}
    & \textsc{C2Rust}                 & 0            & -            & \checkmark   & 268          & -              & -        \\ \cline{2-8}
    & \textsc{C2SaferRust(u)}         & 0.16         & 34.26\%          & \checkmark   & 294          & 3.19           & 11,613   \\ \cline{2-8}
    & \textsc{C2SaferRust(f)}         & 0.13         & 34.78\%          & \checkmark   & 241          & 3.34           & 10,378   \\ \cline{2-8}
    & \textsc{EvoC2Rust}              & \textbf{0.63} & \textbf{68.6\%} & $\times$     & -            & \textbf{1.03}  & \textbf{4,559}  \\ \cline{2-8}
    & LAC2R                           & 0.38         & 67.39\%          & \checkmark   & \textbf{186} & 24.55          & 107,903  \\ \hline \hline

\multirow{5}{*}{\begin{tabular}[c]{@{}c@{}} xzoom \\ (2,524) \end{tabular}}
    & \textsc{C2Rust}                 & 0            & -            & \checkmark   & 99           & -              & -        \\ \cline{2-8}
    & \textsc{C2SaferRust(u)}         & 0.14         & 57.69\%          & \checkmark   & 93           & 2.87           & 9,350    \\ \cline{2-8}
    & \textsc{C2SaferRust(f)}         & 0.14         & 63.64\%          & \checkmark   & 92           & 2.71           & 13,614   \\ \cline{2-8}
    & \textsc{EvoC2Rust}              & \textbf{0.95} & 64.00\%         & $\times$     & -            & \textbf{1.02}  & \textbf{3,072}  \\ \cline{2-8}
    & LAC2R                           & 0.72         & \textbf{100}\%   & \checkmark   & \textbf{64}  & 23.64          & 108,709  \\ \hline \hline

\textbf{Average}
    & \textsc{C2Rust}                 & 0            & -           & \checkmark   & 757.80       & -              & -        \\ \cline{2-8}
    & \textsc{C2SaferRust(u)}         & 0.26         & 56.63\%          & \checkmark   & 960.90       & 2.70           & 7,987    \\ \cline{2-8}
    & \textsc{C2SaferRust(f)}         & 0.29         & 60.67\%          & \checkmark   & 724.40       & 2.79           & 9,339    \\ \cline{2-8}
    & \textsc{EvoC2Rust}              & \textbf{0.75} & 65.8\%          & $\times$     & -            & \textbf{1.07}  & \textbf{4,354}  \\ \cline{2-8}
    & LAC2R                           & 0.54         & \textbf{88.30}\% & \checkmark   & \textbf{541.50} & 19.98       & 86,641   \\ \hline

\end{tabular}
\end{tiny}
\end{table}




\textbf{Comparative Evaluation on small-scale Benchmarks.} Table~\ref{tab:main_exp_tractor} breaks down the results on the TRACTOR benchmark for three subsets of the Public-Tests dataset: \texttt{B01\_organic} (38 cases), \texttt{B01\_synthetic} (85 cases), and \texttt{P00\_perlin\_noise} (1 case). Each covers a different kind of input. \texttt{B01\_organic} includes programs drawn from real codebases that rely on external libraries. \texttt{B01\_synthetic} covers specific features of the C language and the more challenging aspects of translating C. \texttt{P00\_perlin\_noise} is a single larger, project-scale program. In total, these subsets contain 2,620 lines of code for \texttt{B01\_organic}, 3,562 for \texttt{B01\_synthetic}, and 512 for \texttt{P00\_perlin\_noise}, for an overall total of 6,694 lines of code.
For \textsc{C2SaferRust} and \textsc{LAC2R}, we provide TRACTOR test cases during translation, since both methods are built to support test-guided refinement. \textsc{RustAssure} and \textsc{EvoC2Rust} are evaluated using their default pipelines, which do not consume test-vector feedback during translation. \textsc{RustAssure} keeps improving the translations by feeding compiler error messages back to the LLM and, once the code compiles, runs symbolic tests on both versions to ensure behavioral equivalence between the original C and the resulting Rust code. On the other hand, \textsc{EvoC2Rust} fixes errors by relying on the compiler, using its feedback alongside static analysis to generate possible patches and keep adjusting them, without needing pre-written tests to validate changes. For \textsc{SACTOR}, we skip tests during translation to evaluate both library and executable forms the same way, since \textsc{SACTOR}'s verification setup isn't compatible with how \textsc{TRACTOR} handles library testing.
The choice of the LLM used varies across methods and affects performance. \textsc{C2SaferRust}, \textsc{EvoC2Rust}, \textsc{SACTOR}, and \textsc{RustAssure} all rely on GPT-4o, while in our \textsc{LAC2R} work, we leverage both DeepSeek-V3 and GPT-4o.

We first evaluated the original \textsc{C2Rust} translations on the same 124 
programs as a baseline, and we computed $PCR$ and $PPR$ over this full set 
instead of only counting the ones that compiled. These translations achieved 100\% compile and test pass rates under the TRACTOR evaluation harness. 
\textsc{C2SaferRust} and \textsc{LAC2R} matched this on both metrics. This is expected, since both methods start from \textsc{C2Rust} translations 
and refine them incrementally, so they can always fall back on the original 
if a chunk fails.

Table~\ref{tab:main_exp_tractor} demonstrates that \textsc{LAC2R} is the only method that achieves the highest safety ratio (0.95), perfect project-level correctness (100\% $PCR$ and 100\% $PPR$), and the fewest linter warnings (439) on TRACTOR Public-Tests, whereas every other method fails to achieve parity on at least one metric. These results reveal two distinct gaps. First, methods that do not incorporate test-run feedback, namely \textsc{RustAssure} (2.4\% $PPR$) and \textsc{EvoC2Rust} (18\% $PPR$), substantially underperform relative to the perfect correctness attained by \textsc{LAC2R} and \textsc{C2SaferRust}, which both integrate test run feedback. This finding confirms that compiler feedback alone is insufficient to detect semantic discrepancies between a translation that merely compiles and one that is functionally equivalent to the original C code. Second, among the methods that incorporate test-run feedback, \textsc{LAC2R} attains a higher $SR$ (0.95) than \textsc{C2SaferRust} (0.88) while maintaining equivalent correctness, with \textsc{C2SaferRust} also exhibiting inferior code quality. This additional gain in safety and code quality is attributable to \textsc{LAC2R}'s MCTS-based search: the safety-aware reward in Algorithm~\ref{alg:expand} directs tree exploration toward translations with fewer unsafe constructs, thereby surpassing what iterative refinement alone can achieve. The primary tradeoff of \textsc{LAC2R} is inference cost, as it requires the most queries and, together with \textsc{EvoC2Rust}, ranks among the most token intensive methods.

\begin{table}[htbp]
\caption{Comparative results on the TRACTOR Public-Tests. Bold highlights the best performance across the approaches. }
\centering
\label{tab:main_exp_tractor}
\begin{tiny}
\begin{tabular}{|c|c|c|c|c|c|c|c|}
\hline
\begin{tabular}[c]{@{}c@{}} \textit{Dataset} \end{tabular} &
  \multicolumn{1}{c|}{\textit{Methods}} &
  \begin{tabular}[c]{@{}c@{}}SR\end{tabular} & 
  \begin{tabular}[c]{@{}c@{}} PCR \end{tabular}& 
  \begin{tabular}[c]{@{}c@{}} PPR \end{tabular}& 
  \begin{tabular}[c]{@{}c@{}} Linter \\ Warnings \end{tabular}& 
  \begin{tabular}[c]{@{}c@{}} Ave. \\ Queries \end{tabular}& 
  \begin{tabular}[c]{@{}c@{}} Ave. \\ Tokens \end{tabular}  \\ \hline \hline
\multirow{2}{*}{\begin{tabular}[c]{@{}c@{}} TRACTOR Public-Tests \end{tabular}}
    & \textsc{C2Rust}      & 0 & \textbf{100}\% & \textbf{100}\%  & 14,399 & - & - \\ \cline{2-8} 
    & \textsc{C2SaferRust(u)}     & 0.88 & \textbf{100}\% & \textbf{100}\% & 13,727 & \textbf{5.24} & 6,652 \\ \cline{2-8} 
    & \textsc{SACTOR} & 0.8 & 83\% & 63\% & 3,064 & 6.54 & 10,092 \\ \cline{2-8} 
    & \textsc{\textsc{EvoC2Rust}} & 0.04 & 42\% & 18\% & 49,316 & 14.74 & 53,249 \\ \cline{2-8} 
    & \textsc{RustAssure} & 0.72 & 74\% & 2.4\% & 1,026 & 7.86 & \textbf{7,254} \\ \cline{2-8} 
    & LAC2R & \textbf{0.95} & \textbf{100}\% & \textbf{100}\% & \textbf{439} & 25 & 51,788 \\ \hline  
\end{tabular}
\end{tiny}
\end{table}

\textbf{The Effectiveness of Heterogeneous LLMs.} Leveraging heterogeneous LLMs is based on the expectation that diverse code refinements produced by different LLMs have complementary effects leading to improved translation. To evaluate this hypothesis, we compared three variants of LAC2R, such as LAC2R using GPT-4o only, LAC2R using Deepseek V3 only, and the proposed LAC2R using heterogeneous LLMs. Table~\ref{tab:ablation_exp_hetero} shows that LAC2R using heterogeneous LLMs outperforms both variants regardless of which individual LLM is used. It implies that the LLM heterogeneity contributes to consistent performance improvement. 
However, as shown in Table~\ref{tab:ablation_exp_hetero}, the heterogeneous LLMs incurs high cost in terms of LLM tokens and queries.

\begin{table}[htbp]
\caption{The evaluation of LAC2R's heterogeneous LLMs on coreutils. Three variants of LAC2R, such as LAC2R using GPT-4o only, LAC2R using Deepseek V3 only, and the proposed LAC2R using
 heterogeneous LLMs are compared.}
\centering
\label{tab:ablation_exp_hetero}
\begin{tiny}
\begin{tabular}{|c|c|c|c|c|c|c|c|}
\hline
\begin{tabular}[c]{@{}c@{}} \textit{Benchmark} \\ (LoC) \end{tabular} &
  \multicolumn{1}{c|}{\textit{Methods}} &
  \begin{tabular}[c]{@{}c@{}}Safety\end{tabular} &
  \begin{tabular}[c]{@{}c@{}} Ave. \\ Queries \end{tabular}&
  \begin{tabular}[c]{@{}c@{}} Ave. \\ Tokens \end{tabular}&
  \begin{tabular}[c]{@{}c@{}} FRR \end{tabular}&
  \begin{tabular}[c]{@{}c@{}} TPR \end{tabular}&
  \begin{tabular}[c]{@{}c@{}} Linter \\ Warnings \end{tabular} \\ \hline \hline
\multirow{2}{*}{\begin{tabular}[c]{@{}c@{}} split \\ (13848) \end{tabular}}
    & \begin{tabular}[c]{@{}c@{}} LAC2R (GPT-4o) \end{tabular} & 0.29 & 14.78 & 31,168.08 & 86.47\% & 100\% & 293 \\ \cline{2-8}
    & \begin{tabular}[c]{@{}c@{}} LAC2R (Deepseek) \end{tabular} & 0.31 & 20.98 & 40,134.07 & 80.68\% & 100\% & 305 \\ \hline  \hline

\multirow{2}{*}{\begin{tabular}[c]{@{}c@{}} pwd \\ (5859) \end{tabular}}
    & \begin{tabular}[c]{@{}c@{}} LAC2R (GPT-4o) \end{tabular} & 0.75 & 15.38 & 38,000.59 & 94.48\% & 100\% & 98 \\ \cline{2-8}
    & \begin{tabular}[c]{@{}c@{}} LAC2R (Deepseek) \end{tabular} & 0.46 & 16.83 & 33,315.71 & 90.55\% & 100\% & 182 \\ \hline  \hline

\multirow{2}{*}{\begin{tabular}[c]{@{}c@{}} cat \\ (7460) \end{tabular}}
    &  \begin{tabular}[c]{@{}c@{}} LAC2R (GPT-4o) \end{tabular} & 0.44 & 13.92 & 27,796.37 & 93.97\% & 100\% & 173 \\ \cline{2-8}
    & \begin{tabular}[c]{@{}c@{}} LAC2R (Deepseek) \end{tabular} & 0.40 & 14.23 & 25,934.54 & 89.75\% & 100\% & 187 \\ \hline  \hline

\multirow{2}{*}{\begin{tabular}[c]{@{}c@{}} truncate \\ (7181) \end{tabular}}
    &  \begin{tabular}[c]{@{}c@{}} LAC2R (GPT-4o) \end{tabular} & 0.43 & 14.76 & 31,949.39 & 90.32\% & 100\% & 201  \\ \cline{2-8}
    & \begin{tabular}[c]{@{}c@{}} LAC2R (Deepseek) \end{tabular} & 0.25 & 16.88 & 32,916.3 & 90.32\% & 100\% & 192  \\ \hline  \hline

\multirow{2}{*}{\begin{tabular}[c]{@{}c@{}} uniq \\ (8299) \end{tabular}}
    & \begin{tabular}[c]{@{}c@{}} LAC2R (GPT-4o) \end{tabular} & 0.35 & 15.4 & 29,458.98 & 92.21\% & 100\% & 224  \\ \cline{2-8}
    & \begin{tabular}[c]{@{}c@{}} LAC2R (Deepseek) \end{tabular} & 0.33 & 16.61 & 32,671.07 & 89.22\% & 100\% & 238  \\ \hline \hline

\multirow{2}{*}{\begin{tabular}[c]{@{}c@{}} tail \\ (14423) \end{tabular}}
    & \begin{tabular}[c]{@{}c@{}} LAC2R (GPT-4o) \end{tabular} & 0.56 & 15.41 & 42861.54 & 88.34\% & 100\% & 189  \\ \cline{2-8}
    & \begin{tabular}[c]{@{}c@{}} LAC2R (Deepseek) \end{tabular} & 0.48 & 16.25 & 36,371.38 & 88.68\% & 100\% & 261  \\ \hline \hline

\multirow{2}{*}{\begin{tabular}[c]{@{}c@{}} head \\ (8047) \end{tabular}}
    & \begin{tabular}[c]{@{}c@{}} LAC2R (GPT-4o) \end{tabular} &  0.55 & 15.65 & 39,706.32 & 91.50\% & 100\% & 195  \\ \cline{2-8}
    & \begin{tabular}[c]{@{}c@{}} LAC2R (Deepseek) \end{tabular} & 0.50 & 15.77 & 38,414.13 & 86.92\% & 100\% & 190  \\ \hline \hline

\multirow{2}{*}{ \textbf{Average} }
    & \begin{tabular}[c]{@{}c@{}} LAC2R (GPT-4o ) \end{tabular} & 0.4945 & \textbf{15.04} & 34,420.17 & 91.04\% & 100\% & 196.14 \\ \cline{2-8}
    & \begin{tabular}[c]{@{}c@{}} LAC2R (Deepseek) \end{tabular} & 0.3927 & 16.79 & \textbf{34,251.02} &  88.01\% & 100\% & 222.14   \\ \cline{2-8}
    & \begin{tabular}[c]{@{}c@{}} LAC2R (Hetero.) \end{tabular} & \textbf{0.50} & {16.27} & {37,920.17} & \textbf{93.61}\% & 100\% & \textbf{186.57} \\ \hline
\end{tabular}
\end{tiny}
\end{table}

\subsection{Limitations} 

\begin{figure}[hbpt]
    \centering  
    \includegraphics[width=1.0\linewidth,height=3.0cm]{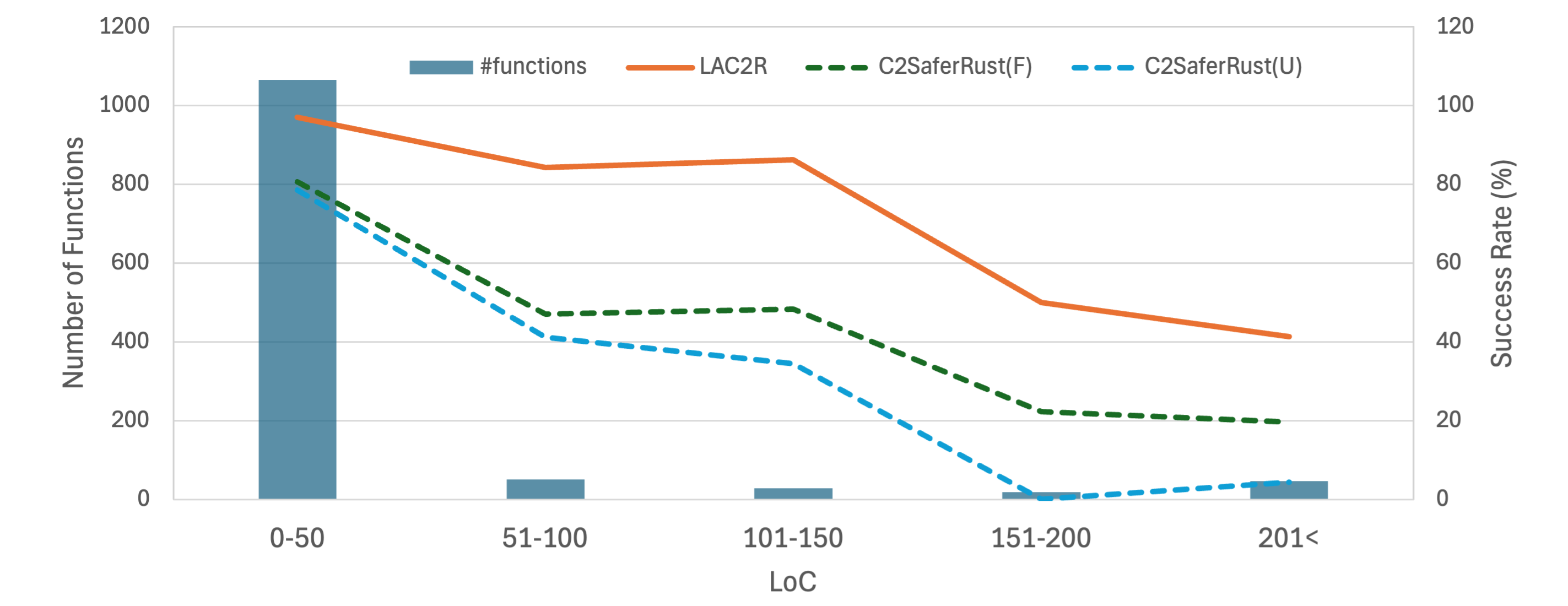} 
    \Description{Decorative image; no additional information.}
    \caption{Function Replacement Rates ($FRR$) across the function-length distribution of the GNU coreutils.}
    \label{fig:loc_success_rates}
\end{figure}

The seven coreutils benchmarks include 1,210 functions, which can be classified using their LoCs. As Figure~\ref{fig:loc_success_rates} shows, most functions are 0-50 LoC long and LAC2R's $FRR$ for the functions is 97\% whereas \textsc{C2SaferRust}'s $FRR$ are around 80\%. As the function length increases, the $FRR$ of all methods decrease. The trend suggests that LLMs struggle to comprehend and translate the entire context of such lengthy functions.

\begin{figure}[hbpt]
    \centering  
    \includegraphics[width=0.8\linewidth,height=2.0cm]{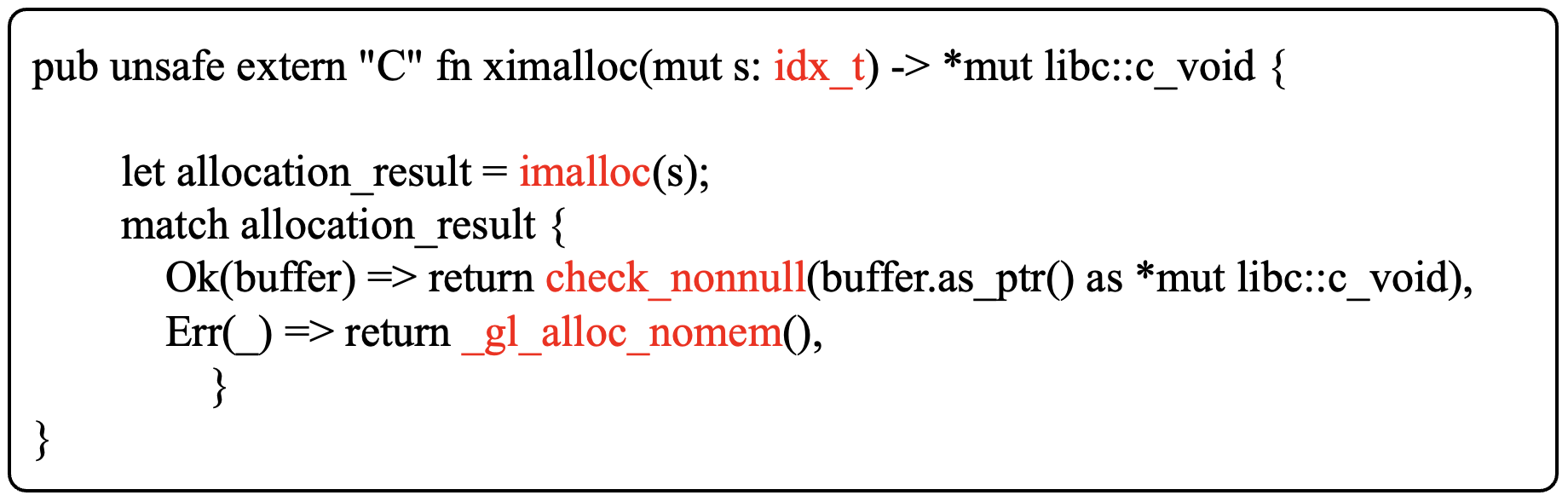} 
    \Description{Decorative image; no additional information.}
    \caption{Failure case 1 of \texttt{tail} on GNU coreutils.}
    \label{fig:fail_case1}
\end{figure}

Figure~\ref{fig:fail_case1} shows an unsafe Rust function \texttt{ximalloc} (<50 LoC) from \texttt{tail}, where LAC2R failed to refine. Its body invokes external functions whose signatures and behavioral contracts are unavailable in the prompt context, preventing LAC2R from determining ownership semantics; MCTS thus fails to find a safe, compilable refinement within the search budget.

\begin{figure}[h]
    \centering  
    \includegraphics[width=0.75\linewidth]{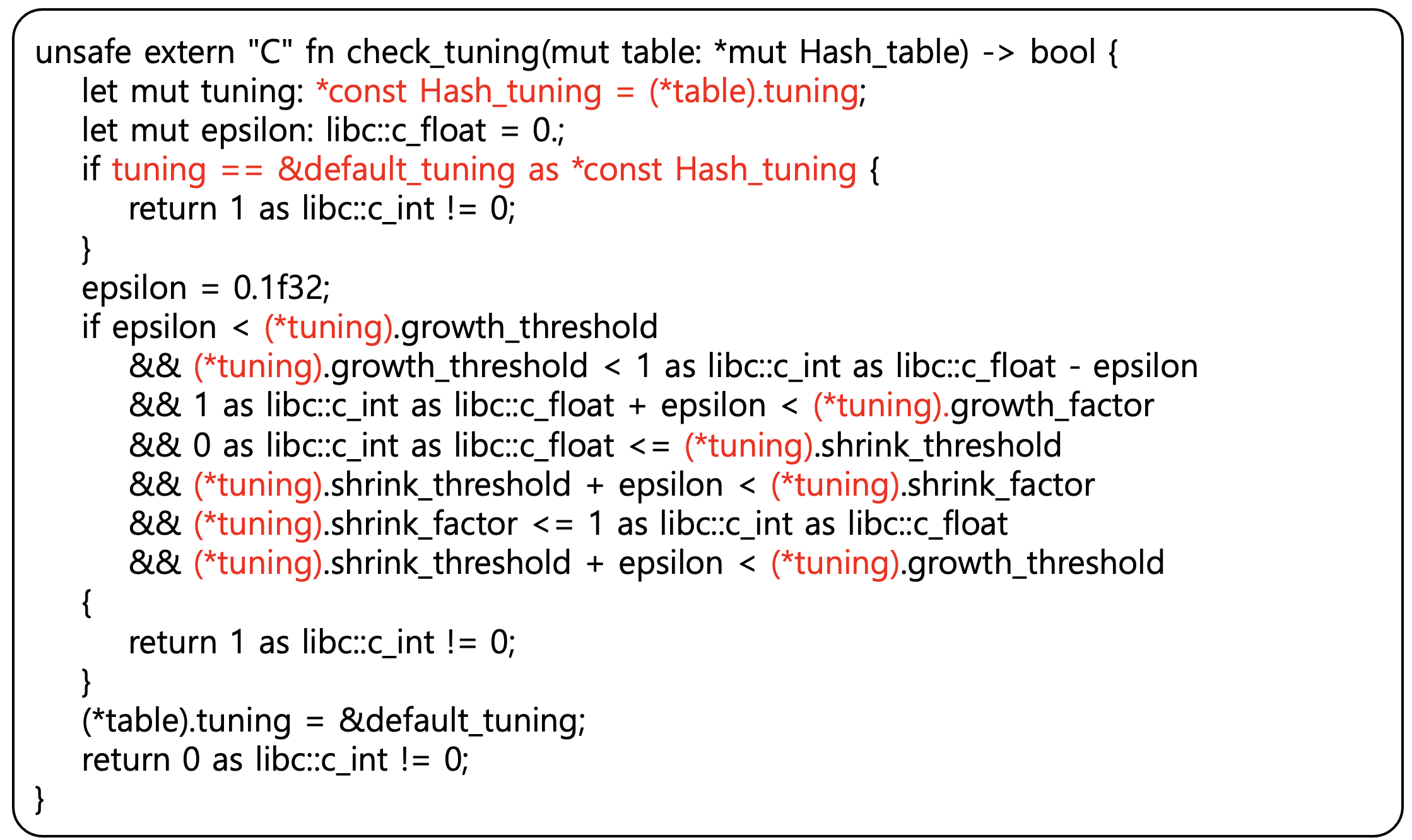} 
    \Description{Decorative image; no additional information.}
    \caption{Failure case 2 of \texttt{tail} on GNU coreutils.}
    \label{fig:fail_case2}
\end{figure}

Figure~\ref{fig:fail_case2} shows a failure where the target function depends on the precise layout and invariants of \texttt{Hash\_table} and \texttt{Hash\_tuning}, whose full definitions were absent from the prompt. Without these structural facts, LAC2R cannot determine field types, aliasing, or lifetimes, and MCTS fails to find a safe refinement. Incorporating richer structural context is an important future direction.

\begin{table*}
  \caption{LLM-based approaches for C-to-Rust translation. Bold highlights the distinct feature of LAC2R.}
  \label{tbl:llm-approaches-structure}
  \begin{footnotesize}
  \centering
  \begin{tabular}{lllll}
    \toprule
    \textit{Name}     & \textit{Translation strategy} & \makecell[l]{\textit{Validator} \\ (\textit{success check})} & \textit{Code analyzer} & \textit{Preprocessor}\\
    \midrule
    \makecell[l]{VERT} & \makecell[l]{Oracle-guided iterative translation}   & \makecell[l]{Bolero~\cite{bolerobook}, Kani~\cite{kanimanual}} & \makecell[l]{Generating an oracle Rust using Wasm} & \makecell[c]{-} \\
    \midrule
    \makecell[l]{\textsc{Flourine}}  & \makecell[l]{Iterative translation validated by cross-language \\ differential fuzzing} & fuzzing-based & Fuzzing  & \makecell[c]{-} \\
    \midrule
    \makecell[l]{\textsc{SpecTra} }     & \makecell[l]{Specification-augmented translation}  & using test cases & \makecell[l]{Generating specification} & \makecell[c]{-}  \\
    \midrule
    \makecell[l]{\textsc{Syzygy} }     & \makecell[l]{Dynamic analysis-guided iterative refinement}  & fuzzing-based & \makecell[l]{Dynamic analysis \\ Fuzzing} & Decomposition \\
    \midrule
    \makecell[l]{\textsc{C2SaferRust} }    &   Iterative refinement of C2Rust output & using test cases & Static analysis & \makecell[l]{Decomposition,  C2Rust}  \\
    \midrule
    \makecell[l]{SACTOR}  & \makecell[l]{ Iterative refinement with two-step translation} & using test cases & Static analysis & Decomposition \\
    \midrule
    \makecell[l]{ SafeTrans }  & \makecell[l]{ Iterative refinement with few-shot learning} & \makecell[l]{using test cases  } & \makecell[c]{-} & \makecell[c]{-} \\
    \midrule
    \makecell[l]{LLMigrate }  & \makecell[l]{Context-aware function-level decomposition \\ and assembly } & \makecell[l]{ Manual test } & Static analysis & \makecell[l]{Decomposition, C2Rust } \\
    \midrule
    \makecell[l]{ \textsc{EvoC2Rust} }  & \makecell[l]{ Skeleton-guided iterative translation } & \makecell[l]{ using test cases} & \makecell[l]{Feature mapping \\ Static analysis} & Skeleton  construction \\
    \midrule
    \makecell[l]{ RustAssure }  & \makecell[l]{ Iterative translation guided by function-level \\ semantic equivalence validation } & \makecell[l]{Differential symbolic testing} & \makecell[l]{-} & \makecell[l]{header file preprocessor \\ struct transpiler \\ struct field usage analyzer} \\
    \midrule
    \makecell[l]{ Rustify }  & \makecell[l]{ Predefined workflow-based multi-agent \\ iterative translation } & \makecell[l]{ using test cases } & \makecell[l]{Static analysis} & \makecell[l]{Decomposition \\ Node merging heuristic} \\
    \midrule
    \makecell[l]{ \textsc{RustMap} }  & \makecell[l]{ Dependency-guided scaffolding \\ for project-scale translation } & \makecell[l]{ using test cases } & \makecell[l]{Static and dynamic analysis} & \makecell[l]{Project-scaffolding \\ C preprocessor} \\
    \midrule
    \makecell[l]{LAC2R (ours)}  & \makecell[l]{ \textbf{MCTS-based structural refinement} } & using test cases & Static analysis & \makecell[l]{Decomposition, C2Rust} \\
    \bottomrule
\end{tabular}
\end{footnotesize}
\end{table*}

\section{Discussion}
\label{sec:discussion}

\noindent \textbf{LLM Cost.} MCTS-based search substantially increases LLM inference costs relative to single-trajectory refinement: LAC2R consumes on average 6--9$\times$ more queries and tokens per function than \textsc{C2SaferRust} (Tables~\ref{tab:main_exp_coreutils} and~\ref{tab:main_exp_laertes}), as each tree expansion requires a full LLM call for code refinement followed by compilation and test validation. Two complementary strategies can mitigate this overhead. At the tree level, heuristic pruning~\cite{DBLP:journals/corr/abs-1911-08265} and rollout-free MCTS~\cite{DBLP:journals/corr/abs-2103-04931} can limit redundant node expansions without degrading solution quality. At the inference level, token-budget-aware reasoning~\cite{han2025tokenbudgetawarellmreasoning} can dynamically cap generation length per call, reducing per-query cost while preserving output quality. Integrating these techniques into LAC2R's MCTS structure is a promising direction for reducing inference overhead while retaining its safety gains.

\noindent \textbf{Correctness Checks.} LAC2R relies on test case-based equivalence checks, consistent with recent approaches~\cite{nitin2025c2saferrusttransformingcprojects,zhou2025llmdrivenmultisteptranslationc}. While convenient and scalable, test-based validation is bounded by the representativeness of the test suite: passing all provided tests does not guarantee semantic equivalence on unseen inputs, as our TRACTOR results confirm (Table~\ref{tab:main_exp_tractor}). Fuzzing-based differential checks~\cite{10.1145/3460319.3464795} can complement this by exploring inputs beyond the fixed test suite, though they depend on harness quality and seed corpus diversity. Symbolic validation~\cite{yang2024vertverifiedequivalentrust} offers stronger path-level guarantees but faces path explosion at the scale of real programs. Incorporating lightweight differential fuzzing into MCTS rollout scoring—penalizing translations that fail on automatically generated inputs—could improve correctness coverage without the scalability penalty of full symbolic verification, and represents a natural extension of LAC2R.

\noindent \textbf{Benchmark Suite.} We evaluate on the seven coreutil and ten Laertes benchmarks, the largest program-level, real-world suites currently used in this domain~\cite{nitin2025c2saferrusttransformingcprojects}, supplemented by TRACTOR Public-Tests as a small-scale benchmark. As Table~\ref{tbl:llm-approaches-experiment} shows, prior works have relied on heterogeneous benchmark sets with limited overlap, making cross-method comparison difficult. CRUST-Bench~\cite{khatry2025crustbenchcomprehensivebenchmarkctosaferust}, a recently introduced dataset of 100 C repositories paired with manually-written Rust function signatures and test cases, is a promising candidate for a unified evaluation platform. Standardizing on such a benchmark and extending coverage to larger programs would enable more rigorous comparison across the rapidly growing set of C-to-Rust translation methods.

\section{Related Work}
\label{sec:related}

Verified Equivalent Rust Transpilation (VERT) combines a rule-based transpilation path with an LLM-generated code candidate~\cite{yang2024vertverifiedequivalentrust}. VERT compiles the C program to WebAssembly and lifts it to Rust using rWasm, producing a semantically correct but often non-idiomatic oracle reference, against which LLM-generated candidates are verified through property-based testing and formal model checking. \textsc{Flourine} proposed cross-language differential fuzzing to compare the I/O behavior of source and translated programs without relying on pre-existing test cases~\cite{eniser2024translatingrealworldcodellms}. \textsc{SpecTra} enhances code translation by incorporating static specifications, input/output test cases, and natural language descriptions individually into the prompt to guide multi-candidate generation~\cite{nitin2024spectraenhancingcodetranslation}. \textsc{Syzygy} combines LLMs with dynamic analysis to extract semantic information such as aliasing behavior and heap allocation sizes for iterative refinement of real-world libraries~\cite{shetty2024syzygydualcodetestc}. \textsc{C2SaferRust} initiates translation from C2Rust output and applies slice-wise iterative refinement driven by static analysis~\cite{nitin2025c2saferrusttransformingcprojects}. SACTOR introduces a two-step pipeline of unidiomatic and idiomatic conversions, where static analysis informs the LLM about pointer semantics and code dependencies in both stages~\cite{zhou2025llmdrivenmultisteptranslationc}. SafeTrans employs few-shot guided repair by supplying the LLM with contextual information and related code snippets, and evaluates whether potential C vulnerabilities are properly resolved in the translated Rust~\cite{farrukh2025safetransllmassistedtranspilationc}. LLMigrate addresses the \textit{laziness} problem—where LLMs omit significant portions of a target code during translation—through function-level decomposition and static analysis~\cite{liu2025llmigratetransforminglazylarge}. \textsc{EvoC2Rust} augments an LLM with safety-preserving mappings between the core linguistic features of C and Rust, constructing a compilable skeleton and incrementally translating individual functions to enable project-scale translation without conventional transpiler preprocessing~\cite{wang2025evoc2rustskeletonguidedframeworkprojectlevel}. \textsc{RustAssure} performs iterative translation guided by function-level semantic equivalence validation using differential symbolic testing, with dedicated preprocessors for header files, structs, and struct field usages~\cite{11334436}. \textsc{Rustify} employs a predefined workflow-based multi-agent framework that iteratively refines translations using static analysis and a node-merging heuristic for function decomposition~\cite{wang2026rustify}. \textsc{RustMap} adopts dependency-guided scaffolding for project-scale translation, leveraging both static and dynamic analysis along with a C preprocessor~\cite{Cai25:Rustmap}.

These existing methods share a similar execution flow at a high-level as illustrated in Figure~\ref{fig:exe_flow}, and all rely on iterative code refinement along a \textit{single} search trajectory. Their detailed structural distinctions are summarized in Table~\ref{tbl:llm-approaches-structure}. In contrast, LAC2R formulates C-to-Rust translation as a sequential decision-making problem and employs MCTS to systematically explore multiple refinement trajectories, enabling principled exploration--exploitation balance via UCT-guided search.

\section{Conclusion}
\label{sec:conclusion}

To address the challenges of C-to-Rust code translation leveraging emerging LLM capabilities, we introduced an MCTS-Guided LLM refinement technique for code translation, LAC2R. These efforts are motivated by the observation that the intermediate steps required for C-to-Rust are not well-defined and there is limited study on how to organize these intermediate steps to construct a correct translation trajectory. LAC2R systematically organizes LLM-generated intermediate steps and improves the possibility of producing a correct translation. Our experimental evaluation on two large-scale, real-world datasets and a small-scale benchmark demonstrates that LAC2R outperforms its counterparts across several metrics, indicating a high likelihood of safe translation.  

\section*{Data-Availability Statement}
The GNU coreutils benchmark originates from the GNU Project 
(\url{https://www.gnu.org/software/coreutils/}). 
The Laertes benchmark suite is publicly available at \url{https://doi.org/10.5281/zenodo.5442253}. 
The TRACTOR Public-Tests benchmark is available at 
\url{https://github.com/DARPA-TRACTOR-Program/PUBLIC-Test-Corpus}. 
The experimental scripts and tool accompanying this paper are available at \url{https://doi.org/10.5281/zenodo.19247427}.


\bibliographystyle{ACM-Reference-Format}
\bibliography{references}

\appendix

\end{document}